\documentclass[a4paper,12pt]{article}

\usepackage{theorem}
\usepackage{amsfonts}
\newfont\fiverm{cmr5}

\usepackage[dvips]{epsfig}	
\input epic.sty
\linethickness{0.4pt}
\def\comment#1{{}}	
\theoremstyle{change}
\newtheorem{thm}{Theorem.\nopagebreak}[section]

\newtheorem{cor}[thm]{Corollary.\nopagebreak}

\newtheorem{prop}[thm]{Proposition.\nopagebreak}

\theorembodyfont{\rmfamily}
\newtheorem{dfn}[thm]{Definition.\nopagebreak}

\newtheorem{conv}[thm]{Convention.\nopagebreak}

\newtheorem{expl}[thm]{Example.\nopagebreak}

\newtheorem{expls}[thm]{Examples.\nopagebreak}

\newtheorem{rem}[thm]{Remark.\nopagebreak}

\newtheorem{alg}[thm]{Algorithm:}

\def\bealg#1{\begin{alg}{\bf #1}\index{Algorithm!#1}\nopagebreak}
\def\ealg{\end{alg}}

\def\Box{{\hbox{\raisebox{0.0em}{\rlap{$\sqcap$}}\kern0em%
            \raisebox{-0.0em}{$\sqcup$}}} } 
\newenvironment{proof}{{\it Proof. }}{~~~\hfill$\Box$\vspace{0.5cm}}
\def\bepf{\begin{proof}}
\def\epf{\end{proof}}

\def\fct#1{\mathop{\rm #1}}	
\def\fns#1{{\mbox{\rm \scriptsize#1}}} 		

\def\cov{\fct{cov}}

\def\dim{\fct{dim}}

\def\forall{~~~\mbox{for all }}

\def\im{\fct{Im}}		

\def\re{\fct{Re}}		

\def\spec{\fct{Spec}}

\def\tr{\fct{tr}}

\def\newl{\hfill\break}				
\def\D{\displaystyle}				

\def\hbar{{\mathchar'26\mkern-9muh}}
\def\kbar{{\mathchar'26\mkern-9muk}}
\def\lp{\mbox{\Large$\,_\urcorner\,$}}   

\def\eps{\varepsilon}
\def\phi{\varphi}
                            
\def\CHI{\mbox{\large $\chi$}}

\def\shalf{\mbox{\small$\frac{1}{2}$\normalsize}}
\def\half{\frac{1}{2}} 
\def\wave{\protect{\footnotesize $\sim$}}

\def\implies{~~~\Rightarrow~~~}

\def\<{\langle} 				
\def\>{\rangle} 				
\def\rangu{\hbox{\raisebox{0.15em}{\rlap{$\sqcap$}}\kern0em%
            \raisebox{-0.19em}{$\sqcup$}} } 

\def\beq{\begin{equation}} 
\def\eeq{\end{equation}} 
\def\lbeq#1{\begin{equation} \label{#1}} 
\def\beqar{\begin{eqnarray}}
\def\eeqar{\end{eqnarray}}
\def\bary{\begin{array}}
\def\eary{\end{array}}
\def\becas{\left\{ \begin{array}{l@{\qquad}l}}
\def\ecas{\end{array} \right.}
\def\benu{\begin{enumerate}}
\def\eenu{\end{enumerate}}
\def\gzit#1{{\rm (\ref{#1})}} 			

\def\Cz{\mathbb{C}}

\def\Ez{\mathbb{E}}

\def\Hz{\mathbb{H}}

\def\Rz{\mathbb{R}}

\def\1{{\mbox{\bf 1}}}

\def\x {{\bf x}}

\makeindex

\begin{document}

\vspace*{-2.5cm}

\begin{center}

{\LARGE \bf Ensembles and experiments}
{\LARGE \bf in classical and quantum physics} \\

\vspace{1.5cm}

\centerline{\sl {\large \bf Arnold Neumaier}}

 \vspace{0.5cm}

\centerline{\sl Institut f\"ur Mathematik, Universit\"at Wien}
\centerline{\sl Strudlhofgasse 4, A-1090 Wien, Austria}
\centerline{\sl email: Arnold.Neumaier@univie.ac.at}
\centerline{\sl WWW: http://www.mat.univie.ac.at/\wave neum/}

\end{center}


{\footnotesize
{\bf Abstract.} 
A philosophically consistent axiomatic approach to classical and 
quantum mechanics is given. The approach realizes a strong 
formal implementation of Bohr's correspondence principle. 
In all instances, classical and quantum concepts are fully parallel: 
the same general theory has a classical realization and a quantum 
realization.
   
Extending the `probability via expectation' 
approach of Whittle to noncommuting quantities,
this paper defines quantities, ensembles, and experiments as 
mathematical concepts and shows how to model complementarity,
uncertainty, probability, nonlocality and dynamics in these terms.
The approach carries no connotation of unlimited repeatability; 
hence it can be applied to unique systems such as the universe.
   
Consistent experiments provide an elegant solution to the reality 
problem, confirming the insistence of the orthodox Copenhagen 
interpretation on that there is nothing but ensembles, 
while avoiding its elusive reality picture. 
The weak law of large numbers explains the emergence of classical 
properties for macroscopic systems.

\vfill
\begin{flushleft}
{\bf Keywords}: 
axiomatization of physics,
Bell inequality,
Bohmian mechanics, 
complementarity,
consistent experiment,
correspondence principle, 
deterministic, 
effect,
elements of physical reality, 
ensemble, 
event, 
expectation, 
flow of truth, 
foundations of quantum mechanics, 
Heisenberg picture, 
hidden variables, 
ideal measurement, 
nonlocality,
foundations of probability,
preparation of states,
quantities, 
quantum correlations,
quantum logic, 
quantum probability,
reference value,
Schr\"odinger picture,
sharpness, 
spin, 
state of the universe, 
uncertainty relation, 
weak law of large numbers,
Young measure
\end{flushleft}

{\bf E-print Archive No.}: quant-ph/0303047
   \newl
{\bf 2002\hspace{.4em} PACS Classification}: 03.65.Bz, 05.30.Ch
   \newl
{\bf 2000\hspace{.4em} MSC Classification}: primary 81P10, 
secondary 81S05

}


\section{Introduction} \label{intro}

\hfill\parbox[t]{8.8cm}{\footnotesize

{\em Do not imagine, any more than I can bring myself to imagine, 
that I should be right in undertaking so great and difficult a task.  
Remembering what I said at first about probability, I will do my best 
to give as probable an explanation as any other -- or rather, more 
probable; and I will first go back to the beginning and try to speak 
of each thing and of all.}

Plato, ca. 367 B.C. \cite{Pla} 
}\nopagebreak

\bigskip 
The purpose of a philosophically consistent axiomatic foundation 
of modern theoretical physicsis is to provide precise mathematical 
concepts which are free of undefined terms and match all
concepts that physicists use to describe their experiments and their
theory, in sufficiently close correspondence to reproduce at least 
that part of physics that is amenable to numerical verification.

\bigskip
This paper is concerned with giving a concise, self-contained 
foundation (more carefully than usual, and without reference to 
measurement) by defining the concepts of quantities, ensembles, 
and experiments, and showing how they give rise to the traditional 
postulates and nonclasssical features of quantum mechanics.

Since it is not clear a priori what it means to `observe' something, 
and since numbers like the fine structure constant or decay rates
can be observed in nature but are only indirectly related to
what is traditionally called an `observable',
we avoid using this notion and employ the more neutral term 
`quantity' to denote quantum operators of interest.

One of the basic premises of this work is that the split between 
classical physics and quantum physics should be as small as possible.
We argue that the differences between classical mechanics and quantum 
mechanics cannot lie in an assumed intrinsic indeterminacy of 
quantum mechanics contrasted to deterministic classical mechanics. 
The only difference between classical mechanics and quantum mechanics 
in the latter's lack of commutativity.
 
Except in the examples, our formalism never distinguishes between the 
classical and the quantum situation. Thus it can be considered as a 
consequent implementation of {\sc Bohr}'s {\em correspondence 
principle}. This also has didactical advantages for teaching: 
Students can be trained to be acquainted at first with the formalism 
by means of intuitive, primarily classical examples. Later, without 
having to unlearn anything, they can apply the same formalism to 
quantum phenomena.

\bigskip
Of course, much of what is done here is based on common wisdom in 
quantum mechanics; see, e.g., {\sc Jammer} \cite{Jam1,Jam2}, 
{\sc Jauch} \cite{Jau}, {\sc Messiah} \cite{Mes}, 
{\sc von Neumann} \cite{vNeu}.
However, apart from being completely rigorous and using no undefined 
terms, the overall setting, the starting points, and the 
interpretation of known results are novel. In particular, the
meaning of the concepts is slightly shifted, carefully crafted and
fixed in a way that minimizes the differences between the classical 
and quantum case.

To motivate the conceptual foundation and to place it into context,
I found it useful to embed the formalism into my philosophy of physics,
while {\em strictly separating the mathematics by using a formal
definition-example-theorem-proof exposition style}. Though I present my 
view generally without using subjunctive formulations or qualifying 
phrases, I do not claim that this is the only way to understand physics.
However, it is an excellent way to understand physics,
integrating different points of view. I believe that my 
philosophical view is fully consistent with the mathematical formalism 
of quantum mechanics and accommodates naturally a number of
puzzling questions about the nature of the world.

\bigskip
The stochastic contents of quantum theory is determined by the 
restrictions noncommutativity places upon the preparation of 
experiments. Since the information going into the preparation is 
always extrapolated from finitely many observations in the past, 
it can only be described in a statistical way, i.e., by ensembles. 

Ensembles are defined by extending to noncommuting quantities
{\sc Whittle}'s \cite{Whi} elegant expectation approach to classical
probability theory. This approach carries no connotation of unlimited 
repeatability; hence it can be applied to unique systems such as the 
universe. The weak law of large numbers relates abstract ensembles and 
concrete mean values over many instances of quantities with the same 
stochastic behavior within a single system.

Precise concepts and traditional results about complementarity, 
uncertainty and nonlocality follow with a minimum of technicalities. 
In particular, nonlocal correlations predicted by {\sc Bell} \cite{Bel} 
and first detected by {\sc Aspect} \cite{Asp} are shown to be already 
consequences of the nature of quantum mechanical ensembles and do not 
depend on hidden variables or on counterfactual reasoning.

The concept of probability itself is derived from that of an ensemble 
by means of a formula motivated from classical ensembles that can be 
described as a finite weighted mean of properties of finitely many 
elementary events.
Probabilities are introduced in a generality supporting so-called
effects, a sort of fuzzy events (related to POV measures that play a 
significant role in measurement theory; see {\sc Busch} et al. 
\cite{BusGL,BusLM}, {\sc Davies} \cite{Dav}, {\sc Peres} \cite{Per3}). 
The weak law of large numbers provides
the relation to the frequency interpretation of probability.
As a special case of the definition, one gets without any effort the
well-known squared probability amplitude formula for transition 
probabilities.

\bigskip 
To separate the conceptual foundations from the thorny issue of how 
the process of performing an experiment affects observations, we
formalize the notion of an experiment by taking into account only
their most obvious aspect, and define
experiments as partial mappings that provide objective 
reference values for certain quantities. Sharpness of quantities is
defined in terms of laws for the reference values; in particular
the squaring law that requires the value of a squared sharp quantity 
$f$ to be equal to the squared value of $f$. It is shown that
the values of sharp quantities must belong to their spectrum, and
that requiring all quantities to be sharp produces contradictions
for Hilbert spaces of dimension $>3$. This is related to well-known 
no-go theorems for hidden variables. (However, recent constructive 
results by {\sc Clifton \& Kent} \cite{CliK} show that in the 
finite-dimensional case there are experiments with a dense set of sharp 
quantities.)

An analysis of a well-known macroscopic reference value, the center
of mass, leads us to reject sharpness as a requirement for consistent
experiments. Considering the statistical foundations of 
thermodynamics, we are instead lead to the view that consistent 
experiments should have the properites of an ensemble. With such 
consistent experiments, the weak law of large numbers explains the 
emergence of classical properties for macroscopic systems.

Quantum reality with reference values defined by consistent 
experiments is as well-behaved and objective as classical macroscopic 
reality with reference values defined by a mass-weighted average 
over constituent values, and lacks sharpness (in the sense of our 
definition) to the same extent as classical macroscopic reality. 
In this interpretation, quantum objects are intrinsically extended 
objects; e.g., the reference radius of a hydrogen atom in the ground 
state is 1.5 times the Bohr radius. 

Thus consistent experiments provide an elegant solution to the reality 
problem, confirming the insistence of the orthodox Copenhagen 
interpretation on that there is nothing but ensembles, while avoiding 
its elusive reality picture. 

Reamrkably, the close analogy between classical and quantum physics
extends even to the deepest level of physics: As shown in 
\cite{Neu.qftev}, classical field theory and quantum field 
theory become almost twin brothers when considered in terms of 
Poisson algebras, which give a common framework for the dynamics of
both classical and quantum systems. (Here we only scratch the surface,
discussing in Section \ref{dynamics} the more foundational aspects
of the dynamics.)

\bigskip
{\bf Acknowledgments.} 
I'd like to thank Waltraud Huyer, Willem de Muynck, Hermann Schichl, 
Tapio Schneider, Victor Stenger, Karl Svozil, Roderich Tumulka and
Edmund Weinmann for useful discussions of earlier versions of this 
manuscript.

\section{Quantities} \label{quantities}

\hfill\parbox[t]{9.5cm}{\footnotesize

{\em But you} [God] 
{\em have arranged all things by measure and number and weight.} 
             
Wisdom of Solomon 11:20, ca. 150 B.C. \cite{Wis}

\bigskip
{\em A quantity in the general sense 
is a property ascribed to phenomena, bodies, or substances that can 
be quantified for, or assigned to, a particular phenomenon, 
body, or substance. [...] 
The value of a physical quantity is the quantitative expression
of a particular physical quantity as the product of a number and a
unit, the number being its numerical value.} 

International System of Units (SI), 1995 \cite{SI}

}\nopagebreak

\bigskip

All our scientific knowledge is based on past observation, and only 
gives rise to conjectures about the future. Mathematical consistency 
requires that our choices are constrained by some formal laws. When we 
want to predict something, the true answer depends on knowledge we do 
not have. We can calculate at best approximations whose accuracy 
can be estimated using statistical techniques (assuming that the 
quality of our models is good).

This implies that we must distinguish between {\em quantities} 
(formal concepts that determine what can possibly be measured or 
calculated) and {\em numbers} (the results of measurements and 
calculations themselves); those quantities that are constant by 
the nature of the concept considered behave just like numbers. 

This terminology is close to the definitions used in the document
defining the international system of units, from which we quoted above.
We deliberately avoid the notion of {\em observables}, since 
it is not clear a priori what it means to `observe' something, 
and since many things (such as the fine structure constant, 
neutrino masses, decay rates, scattering cross sections) 
which can be observed in nature are only indirectly related to
what is traditionally called an `observable'.

Physicists are used to calculating with quantities that they may add 
and multiply without restrictions; if the quantities are complex, the 
complex conjugate can also be formed. It must also be possible to 
compare quantities, at least in certain cases.

Therefore we take as primitive objects of our treatment a set $\Ez$ of 
quantities, such that the sum and the product of quantities is again a 
quantity, and there is an operation generalizing complex conjugation. 
Moreover, we assume that there is an ordering relation that allows us 
to compare two quantities.

Operations on quantities and their comparison are required to satisfy 
a few simple rules; they are called {\bf axioms} since we take them as 
a formal starting point without making any further demands on the
nature of the symbols we are using. Our axioms are motivated by the 
wish to be as general as possible while still preserving the ability 
to manipulate quantities in the manner familiar from matrix algebra. 
(Similar axioms for quantities have been proposed, e.g.,
by {\sc Dirac} \cite{Dir}.)

\begin{dfn} ~

\nopagebreak
(i) $\Ez$ denotes a set whose elements are called {\bf quantities}.
For any two quantities $f,g\in\Ez$, the {\bf sum} $f+g$, the 
{\bf product} $fg$, and the {\bf conjugate} $f^*$ are also quantities.
It is also specified for which pairs of quantities the relation 
$f\geq g$ holds.

The following axioms (Q1)--(Q8) are assumed to hold for all complex 
numbers $\alpha\in\Cz$ and all quantities $f,g,h\in\Ez$.

(Q1) 
~$\Cz \subseteq \Ez$, i.e., complex numbers are special quantities,
where addition, multiplication and conjugation have their traditional
meaning. 

(Q2)
~{$(fg)h=f(gh)$,~~ $\alpha f=f\alpha $,~~ $0f=0$,~~ $1f=f$.}

(Q3)
~{$(f+g)+h=f+(g+h)$,~~ $f(g+h)=fg+fh$,~~ $f+0=f$.}

(Q4)
~{$f^{**}=f$,~~ $(fg)^* =g^* f^* $,~~ $(f+g)^* =f^* +g^*$.}

(Q5)
~{$f^* f =0 \implies f =0$.}

(Q6) 
~$\geq$ is a partial order, i.e., it is reflexive ($f\geq f$),
antisymmetric ($f\geq g \geq f \Rightarrow f=g$) and transitive 
($f\geq g \geq h \Rightarrow f \geq h)$.

(Q7)
~{$f\geq g \implies f+h\geq g+h$.}

(Q8)
~{$f\geq 0 \implies f=f^*$ and $g^*fg\geq 0$.}

(Q9)
~ $1 \geq 0$.

If (Q1)--(Q9) are satisfied we say that $\Ez$ is a {\bf Q-algebra}.

(ii) We introduce the traditional notation
\[ 
f \leq g :\Leftrightarrow g\geq f,
\]
\[
-f:=(-1)f,~~ f-g:=f+(-g), ~~~[f,g]:=fg-gf,
\]
\[
f^0:=1,~~ f^l:=f^{l-1}f~~~ (l=1,2,\dots ),
\]
\[
\re f = \half(f+f^*),~~~\im f = \frac{1}{2i}(f-f^*),
\]
\[
\|f\|=\inf\{\alpha\in\Rz \mid f^*f \leq \alpha^2, \alpha\geq0 \}.
\]
(The infimum of the empty set is taken to be $\infty$.)
$[f,g]$ is called the {\bf commutator} of $f$ and $g$, $\re f$, 
$\im f$ and $\|f\|$ are referred to as the {\bf real part}, the 
{\bf imaginary part}, and the {\bf (spectral) norm} of $f$, 
respectively. The {\bf uniform topology} is the topology induced on
$\Ez$ by declaring a set $E$ open if it contains a ball
$\{f\in\Ez \mid \|f\|<\eps\}$ for some $\eps>0$.

(iii) A quantity $f\in\Ez$ is called {\bf bounded} if 
$\|f\|<\infty$, 
{\bf Hermitian} if $f^*=f$,
and {\bf normal} if $[f,f^*]=0$. More generally, a set $F$ of 
quantities is called {\bf normal} if all its quantities commute with 
each other and with their conjugates.
\end{dfn}

Note that every Hermitian quantity (and in a commutative algebra, 
every quantity) is normal. 

\begin{expls}~

(i) The commutative algebra $\Ez = \Cz^n$ with pointwise multiplication
and componentwise inequalities is a Q-algebra, if vectors with 
constant entries $\alpha$ are identified with $\alpha\in\Cz$.
This Q-algebra describes properties of $n$ classical elementary events;
cf. Example \ref{ex.5.3}(i).

(ii) $\Ez=\Cz^{n\times n}$ is a Q-algebra if complex numbers are 
identified with the scalar multiples of the identity matrix, and
$f\geq g$ iff $f-g$ is Hermitian and positive semidefinite.
This Q-algebra describes quantum systems with $n$ levels. 
For $n=2$, it also describes a single spin, or a qubit.

(iii) The algebra of all complex-valued functions on a set $\Omega$,
with pointwise multiplication and pointwise inequalities is a 
Q-algebra. Suitable subalgebras of such algebras describe classical
probability theory -- cf. Example \ref{ex1.4}(i) -- and classical 
mechanics -- cf. Example \ref{ex.classquant}(i). In the latter case,
$\Omega$ is the phase space of the system considered.

(iv) The algebra of bounded linear operators on a Hilbert space 
$\Hz$, with $f\geq g$ iff $f-g$ is Hermitian and positive semidefinite,
is a Q-algebra. They (or the more general $C^*$-algebras and von 
Neumann algebras) are frequently taken as the basis of nonrelativistic 
quantum mechanics.

(v) The algebra of continuous linear operators on the Schwartz space 
${\cal S}(\Omega_{qu})$ of rapidly decaying functions on a manifold 
$\Omega_{qu}$ is a Q-algebra. It also allows the discussion of 
unbounded quantities. In quantum physics, $\Omega_{qu}$ is the 
configuration space of the system.

Note that physicist generally need to work with unbounded quantities, 
while much of the discussion on foundations takes the more restricted 
Hilbert space point of view. The theory presented here is formulated 
in a way to take care of unbounded quantities, while in our examples, 
we select the point of view as deemed profitable.

\end{expls}

We shall see that, for the general, qualitative aspects of the theory
there is no need to know any details of how to actually perform 
calculations with quantities; this is only needed if one wants to 
calculate specific properties for specific systems. In this respect, 
the situation is quite similar to the traditional axiomatic treatment of
real numbers: The axioms specify the permitted ways to handle formulas 
involving these numbers; and this is enough to derive calculus, say,
without the need to specify either what real numbers {\em are} or 
algorithmic rules for addition, multiplication and division. Of course,
the latter are needed when one wants to do specific calculations but not
while one tries to get insight into a problem. And as the development
of pocket calculators has shown, the capacity for understanding theory 
and that for knowing the best ways of calculation need not even reside 
in the same person.

Note that we assume commutativity only between numbers and quantities.
However, general commutativity of the addition is a consequence of our 
other assumptions. We prove this together with some other useful 
relations. 

\begin{prop}
For all quantities  $f$, $g$, $h\in \Ez$ and $\lambda \in\Cz$,
\lbeq{e.p1}
(f+g)h=fh+gh,~~f-f=0,~~ f+g=g+f
\eeq
\lbeq{e.p2}
[f,f^*]=-2i[\re f,\im f],
\eeq
\lbeq{e.p3}
f^*f\geq 0,~~ ff^*\geq 0.
\eeq
\lbeq{e.p4}
f^*f\leq 0 \implies \|f\|=0 \implies f=0,
\eeq
\lbeq{e.p5}
f\leq g \implies h^*fh\leq h^*gh,~|\lambda|f\leq|\lambda|g,
\eeq
\lbeq{e.p6}
f^*g+g^*f\leq 2\|f\|~\|g\|,
\eeq
\lbeq{e.p7}
\|\lambda f\|=|\lambda| \|f\|,~~~ \|f\pm g\|\leq \|f\|\pm \|g\|,
\eeq
\lbeq{e.p8}
\|f g\|\leq \|f\|~ \|g\|.
\eeq
\end{prop}

\bepf
The right distributive law follows from
\[
\begin{array}{lll}
(f+g)h&=&((f+g)h)^{* *}=(h^* (f+g)^* )^* =(h^* (f^* +g^* ))^* \\
&=&(h^* f^* +h^* g^* )^* =(h^* f^* )^* +(h^* g^* )^* \\
&=&f^{* * }h^{* * }+g^{* * }h^{* * }=fh+gh.
\end{array}
\]
It implies $f-f=1f-1f=(1-1)f=0f=0$. From this, we may deduce that 
addition is commutative, as follows. The quantity $h:=-f+g$
satisfies
\[
-h=(-1)((-1)f+g)=(-1)(-1)f+(-1)g=f-g, 
\]
and we have
\[
f+g=f+(h-h)+g=(f+h)+(-h+g)=(f-f+g)+(f-g+g)=g+f. 
\]
This proves \gzit{e.p1}. If $u=\re f$, $v=\im f$ then $u^*=u,v^*=v$
and $f=u+iv, f^*=u-iv$. Hence 
\[
[f,f^*]=(u+iv)(u-iv)-(u-iv)(u+iv)=2i(vu-uv)=-2i[\re f,\im f],
\]
giving \gzit{e.p2}. \gzit{e.p3}--\gzit{e.p5} follow directly from 
(Q7) -- (Q9). Now let $\alpha=\|f\|$, $\beta=\|g\|$. Then 
$f^*f\leq \alpha^2$ and $g^*g\leq \beta^2$. Since
\[
\begin{array}{lll}
0\leq (\beta f - \alpha g)^*(\beta f - \alpha g)&=&
\beta^2f^*f-\alpha\beta(f^*g+g^*f)+\alpha^2 g^*g\\
&\leq& \beta^2\alpha^2 \pm\alpha\beta(f^*g+g^*f) +\alpha^2 g^*g,
\end{array}
\] 
$f^*g+g^*f\leq 2\alpha\beta$ if $\alpha\beta\neq 0$, and for
$\alpha\beta=0$, the same follows from \gzit{e.p4}. Therefore
\gzit{e.p6} holds. The first half of \gzit{e.p7} is trivial, and
the second half follows for the plus sign from 
\[
(f+g)^*(f+g)=f^*f+f^*g+g^*f+g^*g
\leq \alpha^2+ 2\alpha\beta+\beta^2=(\alpha+\beta)^2,
\]
and then for the minus sign from the first half.
Finally, by \gzit{e.p5},
\[
(fg)^*(fg)=g^*f^*fg\leq g^*\alpha^2g=\alpha^2g^*g\leq\alpha^2\beta^2.
\]
This implies \gzit{e.p8}.
\epf

\begin{cor}\label{c1.3}~

(i) Among the complex numbers, precisely the nonnegative real numbers
$\lambda$ satisfy $\lambda\geq 0$.

(ii) For all $f\in\Ez$, $\re f$ and $\im f$ are Hermitian. $f$ is 
Hermitian iff $f=\re f$ iff $\im f=0$. If $f,g$ are commuting 
Hermitian quantities then $fg$ is Hermitian, too.

(iii) $f$ is normal iff $[\re f,\im f]=0$.
\end{cor}

\bepf
(i) If $\lambda$ is a nonnegative real number then $\lambda=f^*f\geq0$ 
with $f=\sqrt{\lambda}$. If $\lambda$ is a negative real number then 
$\lambda=-f^*f\leq0$ with $f=\sqrt{-\lambda}$, and by antisymmetry,
$\lambda\geq0$ is impossible. If $\lambda$ is a nonreal number then 
$\lambda\neq\lambda^*$ and $\lambda\geq0$ is impossible by (Q8).

The first two assertions of (ii) are trivial, and the third holds since
$(fg)^*=g^*f^*=gf=fg$ if $f,g$ are Hermitian and commute.

(iii) follows from \gzit{e.p2}. 
\epf

Thus, in conventional terminology (see, e.g., {\sc Rickart} \cite{Ric}),
$\Ez$ is a {\bf partially ordered nondegenerate *-algebra with unity},
but not necessarily with a commutative multiplication. 

\begin{rem} 
In the realizations of the axioms I know of, e.g., in $C^*$-algebras 
({\sc Rickart} \cite{Ric}), we also have the relations
\[
\|f^*\|=\|f\|,~~~\|f^*f\|=\|f\|^2,
\]
and
\[
0\leq f \leq g \implies f^2 \leq g^2,
\]
but I have not been able to prove these from the present axioms,
and they were not needed to develop the theory.
\end{rem}

As the example
$\Ez=\Cz^{n\times n}$ shows, $\Ez$ may have zero
divisors, and not every nonzero quantity need have an inverse.
Therefore, in the manipulation of formulas, precisely the same 
precautions must be taken as in ordinary matrix algebra.

\section{Complementarity} \label{compl}

\hfill\parbox[t]{6.8cm}{\footnotesize

{\em You cannot have the penny and the cake.}

Proverb

}\nopagebreak

\bigskip
The lack of commutativity gives rise to the phenomenon of
complementarity, expressed by inequalities that demonstrate the danger 
of simply thinking of quantities in terms of numbers. 

\begin{dfn}
Two Hermitian quantities $f,g$ are called {\bf complementary} if there
is a real number $\gamma>0$ such that
\lbeq{e.compl}
(f-x)^2+(g-y)^2 \geq \gamma^2~~~\mbox{for all }x,y\in\Rz.
\eeq
\end{dfn}

Complementarity captures the phenomenon where two quantities do 
not have simultaneous sharp classical `values'. 

\begin{thm}~

(i) In $\Cz^{n\times n}$, two complementary quantities cannot commute.

(ii) A (commutative) Q-algebra of complex-valued functions on a 
set $\Omega$ contains no complementary pair of quantities.
\end{thm}

\bepf
(i) Any two commuting quantities $f,g$ have a common eigenvector $\psi$.
If $f\psi=x\psi$ and $g\psi=y\psi$ then 
$\psi^*((f-x)^2+(g-y)^2)\psi=0$, whereas \gzit{e.compl} implies 
\[
\psi^*(f-x)^2+(g-y)^2)\psi\geq \gamma^2\psi^*\psi >0.
\] 
Thus $f,g$ cannot be complementary.

(ii) Setting $x=f(\omega)$, $y=g(\omega)$ in \gzit{e.compl}, we find 
$0\geq \gamma^2$, contradicting $\gamma>0$.
\epf

I have not been able to decide whether a commutative Q-algebra 
containing complementary quantities exist, or whether complementary 
quantities in an infinite-dimensional Q-algebra can possibly commute. 
(It is impossible when there is a joint spectral resolution.)

\begin{expls}~

(i) $\Cz^{2\times2}$ contains a complementary pair of quantities.
For example, the Pauli matrices
\lbeq{e.pauli}
\sigma_1 =\left(\begin{array}{l}0~~1\\1~~0\end{array}\right),~~
\sigma_3 =\left(\begin{array}{l}1~~\phantom{-}0\\0~~-1
\end{array}\right)
\eeq
are complementary; see Proposition \ref{p.comp}(i) below.

(ii) The algebra of bounded linear operators on a Hilbert space of 
dimension greater than one contains a complementary pair of 
quantities, since it contains a subalgebra isomorphic to 
$\Cz^{2\times2}$.

(iii) In the algebra of all linear operators on the Schwartz space 
${\cal S}(\Rz)$, {\bf position} $q$, defined by 
\[
(qf)(x)=xf(x),
\]
and {\bf momentum} $p$, defined by 
\[
(pf)(x)=-i\hbar f'(x),
\]
where $\hbar>0$ is Planck's constant, are complementary.
Since $q$ and $p$ are Hermitian, this follows from the easily
verified {\bf canonical commutation relation}
\lbeq{ccr}
[q,p]=i\hbar
\eeq
and Proposition \ref{p.comp}(ii) below.

The name 'complementarity' comes from the fact that if one
finds in an experiment (reasonably) sharp values for position, 
one gets the 'particle view' of quantum mechanics, while if one finds 
(reasonably) sharp values for momentum, 
one gets the 'wave view'. The views are complementary in the sense 
that while, correctly interpreted (namely as the position and momentum 
representation, respectively), both descriptions are formally 
equivalent, nevertheless arbitrarily sharp values for both position 
and momentum cannot be realized simultaneously in experiments.
See Section \ref{uncertainty}, in particular the discussion after 
Proposition \ref{p5.1}, for lower bounds on the uncertainty, 
and Section \ref{ideal} for the concept of (idealized) sharpness.

\end{expls}

\begin{prop}\label{p.comp}~

(i) The Pauli matrices \gzit{e.pauli} satisfy
\lbeq{e6.uncpauli}
(\sigma_1-s_1)^2+(\sigma_3-s_3)^2 \geq 1 
~~~\mbox{for all } s_1,s_3\in\Rz.
\eeq

(ii) Let $p,q$ be Hermitian quantities satisfying $[q,p]=i\hbar$.
Then, for any $k,x\in\Rz$ and any positive $\Delta p,\Delta q \in\Rz$,
\lbeq{e6.unc}
\Big(\frac{p-k}{\Delta p}\Big)^2+\Big(\frac{q-x}{\Delta q}\Big)^2
\geq \frac{\hbar}{\Delta p \Delta q}.
\eeq
\end{prop}
\begin{proof}
(i) A simple calculation gives
\[
(\sigma_1-s_1)^2+(\sigma_3-s_3)^2-1=\left(\begin{array}{cc}
s_1^2+(1-s_3)^2 & -2s_1 \\
-2s_1           & s_1^2+(1+s_3)^2 \\
\end{array}\right) \geq 0,
\]
since the diagonal is nonnegative and the determinant is
$(s_1^2+s_3^2-1)^2\geq 0$.

(ii) The quantities $f=(q-x)/\Delta q$ and $g=(p-k)/\Delta p$ are 
Hermitian and satisfy $[f,g]=[q,p]/\Delta q\Delta p=i\kappa$ where 
$\kappa=\hbar/\Delta q\Delta p$. Now \gzit{e6.unc} follows from
\[
0\leq (f+ig)^*(f+ig)=f^2+g^2+i[f,g]=f^2+g^2-\kappa.
\]
\end{proof}

The complementarity of position and momentum expressed by \gzit{e6.unc}
is the deeper reason for the Heisenberg uncertainty relation discussed 
later in \gzit{e6.unc0} and \gzit{e6.unc1}.

\section{Ensembles} 
\label{ensembles}

\hfill\parbox[t]{8.8cm}{\footnotesize

{\em We may assume that words are akin to the matter which they 
describe; when they relate to the lasting and permanent and 
intelligible, they ought to be lasting and unalterable, and, as far 
as their nature allows, irrefutable and immovable -- nothing less.  
But when they express only the copy or likeness and not the eternal 
things themselves, they need only be likely and analogous to the real 
words. As being is to becoming, so is truth to belief.}

Plato, ca. 367 B.C. \cite{Pla}
}\nopagebreak

\bigskip
The stochastic nature of quantum mechanics is usually discussed in
terms of {\em probabilities}. However, from a strictly logical point 
of view, this has the drawback that one gets into conflict with the 
traditional foundation of probability theory by 
{\sc Kolmogorov} \cite{Kol}, which does not extend to the 
noncommutative case. Mathematical physicists (see, e.g., 
{\sc Parthasarathy} \cite{Par}, {\sc Meyer} \cite{Mey}) developed a 
far reaching quantum probability calculus based on Hilbert space 
theory. But their approach is highly formal, drawing its motivation 
from analogies to the classical case rather than from the common 
operational meaning.

{\sc Whittle} \cite{Whi} presents a much less known but very elegant 
alternative approach to classical probability theory, equivalent to 
that of Kolmogorov, that treats {\em expectation} as the basic concept 
and derives probability from axioms for the expectation. (See the 
discussion in \cite[Section 3.4]{Whi} why, for historical reasons, 
this has so far remained a minority approach.)  

The approach via expectations is easy to motivate, leads quickly to 
interesting results, and extends without trouble to the quantum 
world, yielding the ensembles (`mixed states') of traditional quantum 
physics. As we shall see, explicit probabilities enter only at 
a very late stage of the development.

A significant advantage of the expectation approach compared with the 
probability approach is that it is intuitively more removed from a 
connotation of `unlimited repeatability'. Hence it can be naturally 
used for {\em unique} systems such as the set of all natural 
globular proteins (cf., e.g., {\sc Neumaier} \cite{Neu.prot}), the 
climate of the earth, or the universe, and to deterministic, 
pseudo-random behavior such as rounding errors in floating point 
computations (cf., e.g., {\sc Higham} \cite[Section 2.6]{Hig}), once 
these have enough complexity to exhibit finite {\em internal} 
repetitivity to which the weak law of large numbers 
(Theorem \ref{t.weaklaw} below) may be applied.

\bigskip
The axioms we shall require for meaningful expectations are those
trivially satisfied for weighted averages of a finite ensemble of 
observations. While this motivates the form of the axioms and the
name `ensemble' attached to the concept, there is no need at all to 
interpret expectation as an average (or, indeed, the 'ensemble' as
a multitude of actual or possible 'realizations'); 
this is appropriate only in certain classical situations.
 
In general, ensembles are simply a way to consistently organize 
structured data obtained by some process of observation. 
For the purpose of statistical analysis and prediction, it is 
completely irrelevant what this process of observation entails. 
What matters is only that for certain quantities observed values are 
available that can be compared with their expectations. 
The expectation of a quantity $f$ is simply a value near which, based 
on the theory, we should expect an observed value for $f$. At the same 
time, the standard deviation serves as a measure of the amount to 
which we should expect this nearness to deviate from exactness. 
(For more on observed values, 
see Sections \ref{ideal}--\ref{consistent}.)

\begin{dfn}~

\nopagebreak
(i) An {\bf ensemble} is a mapping $^-$ that assigns to each quantity 
$f \in \Ez$ its {\bf expectation} $\overline{f}=:\< f\> \in \Cz$ 
such that for all $f,g \in \Ez$, $\alpha \in \Cz$,

(E1)~ $\<1\> =1, ~~\<f^*\>=\<f\>^*,~~ \< f+g\> =\<f\> +\<g\> $, 

(E2)~ $\<\alpha f\> =\alpha\<f\>$, 

(E3)~ If $f \ge 0$ then $\<f\> \ge 0$,

(E4)~ If $f_l\in\Ez,~ f_l \downarrow 0$ then $\inf \<f_l\> = 0$.

Here $f_l \downarrow 0$ means that the $f_l$ converge almost
everywhere to $0$, and $f_{l+1}\leq f_l$ for all $l$.

(ii) The number
\[
\cov(f,g):=\re \<(f-\overline{f})^*(g-\overline{g}) \>
\]
is called the {\bf covariance} of $f,g\in\Ez$. Two quantities $f,g$ are 
called {\bf correlated} if $\cov(f,g)\neq0$, and {\bf uncorrelated} 
otherwise.

(iii) The number 
\[
\sigma(f):=\sqrt{\cov(f,f)}
\]
is called the {\bf uncertainty} or {\bf standard deviation} of 
$f\in\Ez$ in the ensemble $\<\cdot\>$. 

\end{dfn}

This definition generalizes the expectation axioms of
{\sc Whittle} \cite[Section 2.2]{Whi} for classical probability theory
and the definitions of elementary classical statistics.
Note that (E3) ensures that $\sigma(f)$ is a nonnegative real number
that vanishes if $f$ is a constant quantity (i.e., a complex number). 

(We shall not use axiom (E4) in this paper and therefore do not go
into technicalities about almost everywhere convergence, which are 
needed to get equivalence to Kolmogorov's probability theory in the 
classical case.)

\begin{expls}  \label{ex.5.3}~

(i) {\bf Finite probability theory.}
In the commutative Q-algebra $\Ez = \Cz^n$ with componentwise 
multiplication and componentwise inequalities, every linear functional 
on $\Ez$, and in particular every ensemble, has the form 
\lbeq{e5.fin}
\<f\>=\sum_{k=1}^n p_k f_k
\eeq
for certain weights $p_k$. The ensemble axioms hold precisely 
when the $p_k$ are nonnegative and add up to one; thus $\<f\>$ 
is a weighted average, and the weights have the intuitive meaning of 
`probabilities'. 

Note that the weights can be recovered from the expectation
by means of the formula $p_k=\<e_k\>$, where $e_k$ is the unit vector
with a one in component $k$.

(ii) {\bf Quantum mechanical ensembles.}
In the Q-algebra $\Ez$ of bounded linear operators on a Hilbert space 
$\Hz$, quantum mechanics describes a {\bf pure ensemble} 
(traditionally called a `pure state') 
by the expectation
\[
\<f\>:=\psi^*f\psi,
\]
where $\psi\in\Hz$ is a unit vector. And quantum thermodynamics 
describes an {\bf equilibrium ensemble} by the expectation
\[
\<f\>:=\tr e^{-S/\kbar}f, 
\]
where $\kbar>0$ is the {\bf Boltzmann constant}, and $S$ is a Hermitian
quantity with $\tr e^{-S/\kbar}=1$ called the {\bf entropy} whose 
spectrum is discrete and bounded below.
In both cases, the ensemble axioms are easily verified.

\end{expls}

\begin{prop} \label{p5.2}

For any ensemble,

(i) $f\leq g \implies \<f\> \leq \<g\>$.

(ii) For $f,g\in\Ez$,
\[
\cov(f,g)=\re(\<f^*g\>-\<f\>^*\<g\>),
\]
\[
\<f^*f\>=\<f\>^*\<f\>+\sigma(f)^2,
\]
\[
|\<f\>|\leq\sqrt{\<f^*f\>}.
\]

(iii) If $f$ is Hermitian then $\bar f = \<f\>$ is real and
\[
\sigma(f)=\sqrt{\<(f-\overline{f})^2 \>}
=\sqrt{\<f^2\>-\<f\>^2}.
\]

(iv) Two commuting Hermitian quantities $f,g$ are uncorrelated iff
\[
\<fg\>=\<f\>\<g\>.
\]

\end{prop}

\bepf
(i) follows from (E1) and (E3).

(ii) The first formula holds since
\[
\<(f-\bar f)^*(g-\bar g)\>
=\<f^*g\>-\bar f^*\<g\>-\<f\>^*\bar g +\bar f^*\bar g 
= \<f^*g\>-\<f\>^*\<g\>.
\]
The second formula follows for $g=f$, using (E1), and the third 
formula is an immediate consequence.

(iii) follows from (E1) and (ii).

(iv) If $f,g$ are Hermitian and commute the $fg$ is Hermitian by 
Corollary \ref{c1.3}(ii), hence $\<fg\>$ is real. By (ii),
$\cov(f,g)=\<fg\>-\<f\>\<g\>$, and the assertion follows.

\epf

Fundamental for the practical use of ensembles, and basic to 
statistical mechanics, is the {\bf weak law of large numbers}:

\begin{thm}\label{t.weaklaw} 
For a family of quantities $f_l$ $(l=1, \ldots , N)$ with
constant expectation $\< f_l \> = \mu$, the {\bf mean value}
\[
  \bar f := \frac{1}{N} \D \sum ^N _{l=1} f_l
\]
satisfies
\[
  \< \bar f \> =\mu.
\]
If, in addition, the $f_l$ are uncorrelated and have constant standard 
deviation $\sigma(f_l)=\sigma$ then
\lbeq{e.sigN}
\sigma (\bar f) = \sigma/\sqrt{N}
\eeq
becomes arbitrarily small as $N$ becomes sufficiently large. 
\end{thm}

\bepf 
We have
\[
\<\bar f\> =\frac{1}{N}(\<f_1\>+\dots+\<f_N\> )
=\frac{1}{N}(\mu+\dots+\mu)=\mu
\] 
and
\[
\bar f^*\bar f=\frac{1}{N^2}\Big(\sum_jf_j\Big)^*\Big(\sum_kf_k\Big)
=N^{-2}\sum_{j,k}f_j^*f_k.
\]
Now
\[
\<f_j^*f_j\>=\<f_j\>^*\<f_j\>+\sigma(f_j)^2=|\mu|^2+\sigma^2
\]
and, if the $f_l$ are uncorrelated, for $j\neq k$,
\[
\<f_j^*f_k+f_k^*f_j\>=2\re \<f_j^*f_k\>
=2\re \<f_j\>^*\<f_k\>=2\re \mu^*\mu=2|\mu|^2.
\]
Hence 
\[
\begin{array}{lll}
\sigma(\bar f)^2 &=& \<\bar f^*\bar f\>-\<\bar f\>^*\<\bar f\> \\
&=& N^{-2}\Big(N(\sigma^2+|\mu|^2)+{N \choose 2}2|\mu|^2\Big)-\mu^*\mu
=N^{-1}\sigma^2,
\end{array}
\]
and the assertions follow.
\epf

As a significant body of work in probability theory shows, the
conditions under which $\sigma(\bar f)\to 0$ as $N\to\infty$ can
be significantly relaxed.

\section{Uncertainty} 
\label{uncertainty}

\hfill\parbox[t]{8.8cm}{\footnotesize

{\em For you do not know which will succeed, whether this or that, or 
whether both will do equally well.}

Kohelet, ca. 250 B.C. \cite{Koh}

\bigskip
{\em But if we have food and clothing, we will be content with that.}

St. Paul, ca. 60 A.D. \cite{Pau3} 
}\nopagebreak

\bigskip
Due to our inability to 
prepare experiments with a sufficient degree of sharpness to know 
with certainty everything about a system we investigate,
we need to describe the preparation of experiments in a 
stochastic language that permits the discussion of such uncertainties; 
in other words, we shall model prepared experiments by ensembles.

Formally, the essential difference between classical mechanics 
and quantum mechanics in the latter's lack of commutativity.
While in classical mechanics there is in principle no lower
limit to the uncertainties with which we can prepare the quantities
in a system of interest,
the quantum mechanical uncertainty relation for noncommuting 
quantities puts strict limits on the uncertainties in the preparation
of microscopic ensembles. Here, {\em preparation} is defined informally 
as bringing the system into an ensemble such that measuring certain 
quantities gives values that agree with the expectation to an accuracy 
specified by given uncertainties.

In this section, we discuss the limits of the accuracy to which this 
can be done.

\begin{prop} \label{p5.1}~\nopagebreak

(i) The {\bf Cauchy--Schwarz inequality}  
\[
|\< f^*g \>|^2 \le \< f^*f \>\< g^*g \>
\]
holds for all $f,g\in\Ez$.

(ii) The {\bf uncertainty relation}
\[
\sigma(f)^2\sigma(g)^2 
\geq |\cov(f,g)|^2+\left|\shalf\<f^*g-g^*f\>\right|^2
\]
holds for all $f,g\in\Ez$.

(iii) For $f,g\in\Ez$, 
\lbeq{ecov1}
\cov(f,g)=\cov(g,f)=\shalf(\sigma(f+g)^2-\sigma(f)^2-\sigma(g)^2),
\eeq
\lbeq{ecov}
|\cov(f,g)| \leq \sigma(f)\sigma(g), 
\eeq
\lbeq{esig}
\sigma(f+g) \leq \sigma(f)+\sigma(g).
\eeq
In particular,
\lbeq{e.prodbound}
|\<fg\>-\<f\>\<g\>|\leq\sigma(f)\sigma(g) 
~~~\mbox{for commuting Hermitian } f,g. 
\eeq

\end{prop}

\bepf
(i) For arbitrary $\alpha ,\beta\in \Cz$ we have
\[
\begin{array}{ll}
0&\le \<(\alpha f-\beta g)^*(\alpha f-\beta g )\> \\
&=\alpha ^* \alpha \< f^*f \>-\alpha ^* \beta \< f^*g \>
-\beta ^*\alpha \< g^*f \>+\beta\beta^* \< g^*g \>\\
&=|\alpha |^2\< f^*f \>-2\re(\alpha ^* \beta \< f^*g \>)
+|\beta|^2\< g^*g \>
\end{array}
\]
We now choose $\beta=\< f^*g \>$, and obtain for arbitrary
real $\alpha $ the inequality
\lbeq{f.8}
0\le \alpha ^2\< f^*f \>
-2\alpha |\< f^*g \>|^2+|\< f^*g \>|^2\< g^*g \>.
\eeq
The further choice $\alpha=\< g^*g \>$ gives
\[
0\le \< g^*g \>^2\< f^*f \>-\< g^*g \>|\< f^*g \>|^2.
\]
If $\< g^*g \>>0$, we find after division by $\< g^*g \>$ that (i) 
holds. And if $\< g^*g \>\le 0$ then $\< g^*g \>=0$ and we have 
$\< f^*g \>=0$ since otherwise a tiny $\alpha $ produces a negative
right hand side in \gzit{f.8}. Thus (i) also holds in this case.

(ii) Since $(f-\bar f)^*(g-\bar g)-(g-\bar g)^*(f-\bar f)=f^*g-g^*f$,
it is sufficient to prove the uncertainty relation for the case of
quantities $f,g$ whose expectation vanishes. In this case, (i) implies
\[
(\re \<f^*g\>)^2 +(\im \<f^*g\>)^2 =|\<f^*g\>|^2 \leq 
\< f^*f \>\< g^*g \> = \sigma(f)^2\sigma(g)^2.
\]
The assertion follows since $\re \<f^*g\>=\cov(f,g)$ and
\[
i\im \<f^*g\>=\shalf(\<f^*g\>-\<f^*g\>^*)=\shalf\<f^*g-g^*f\>.
\]

(iii) Again, it is sufficient to consider the case of
quantities $f,g$ whose expectation vanishes. Then
\lbeq{esig1}
\begin{array}{lll}
\sigma(f+g)^2 &=& \<(f+g)^*(f+g)\>
=\<f^*f\>+\<f^*g+g^*f\>+\<g^*g\>\\
&=& \sigma(f)^2+2\cov(f,g)+\sigma(g)^2,
\end{array}
\eeq
and \gzit{ecov1} follows. \gzit{ecov} is an immediate consequence of
(ii), and \gzit{esig} follows easily from \gzit{esig1} and 
\gzit{ecov}. Finally, \gzit{e.prodbound} is a consequence of 
\gzit{ecov} and Proposition \ref{p5.2}(iii).
\epf

In the classical case of commuting Hermitian quantities, the 
uncertainty relation just reduces to the well-known inequality 
\gzit{ecov} of classical statistics. For noncommuting Hermitian 
quantities, the uncertainty relation is stronger. In particular, we may
deduce from the commutation relation \gzit{ccr} for position $q$ and 
momentum $p$ {\sc Heisenberg}'s \cite{Hei,Rob} uncertainty relation
\lbeq{e6.unc0}
\sigma(q)\sigma(p)\geq \shalf\hbar.
\eeq
Thus {\em no ensemble exists 
where both $p$ and $q$ have arbitrarily small standard deviation}. 
(More general noncommuting Hermitian quantities $f,g$ may have 
{\em some} ensembles with $\sigma(f)=\sigma(g)=0$, namely among those 
with $\<fg\>=\<gf\>$.)

Putting $k=\bar p$ and $x=\bar q$, taking expectations in 
\gzit{e6.unc} and using Proposition \ref{p5.2}(iii), we find another 
version of the uncertainty relation, implying again that $\sigma(p)$ 
and $\sigma(q)$ cannot be made simultaneously very small:
\lbeq{e6.unc1}
\Big(\frac{\sigma(p)}{\Delta p}\Big)^2
+\Big(\frac{\sigma(q)}{\Delta q}\Big)^2
\geq\frac{\hbar}{\Delta p \Delta q}.
\eeq
Heisenberg's relation \gzit{e6.unc0} follows from it by putting
$\Delta p = \sigma(p)$ and $\Delta q = \sigma(q)$.

The same argument shows that no ensemble exists 
where two complementary quantities both have arbitrarily small 
standard deviation.
(More general noncommuting Hermitian quantities $f,g$ may have 
{\em some} ensembles with $\sigma(f)=\sigma(g)=0$, namely among those 
with $\<fg\>=\<gf\>$.)

We now derive a characterization of the quantities $f$ with vanishing
uncertainty, $\sigma(f)=0$; in classical probability theory these 
correspond to quantities (random variables) that have fixed values 
in every realization.

\begin{dfn}
We say a quantity $f$ {\bf vanishes} in the ensemble $\<\cdot\>$ if
\[
\<f^*f\>=0.
\]
\end{dfn}

\begin{thm}~

(i) $\sigma(f)=0$ iff $f-\<f\>$ vanishes.

(ii) If $f$ vanishes in the ensemble $\<\cdot\>$ then $\<f\>=0$.

(iii) The set $V$ of vanishing quantities satisfies
\[
f+g\in V~~~\mbox{if } f,g\in V,
\]
\[
fg\in V~~~\mbox{if $g\in V$ and $f\in \Ez$ is bounded},
\]
\[
f^2\in V~~~\mbox{if $f\in V$ is Hermitian}.
\]
\end{thm}

\bepf
(i) holds since $g=f-\<f\>$ satisfies $\<g^*g\>=\sigma(f)^2$.

(ii) follows from Proposition \ref{p5.2}(ii).

(iii) If $f,g\in V$ then $\<f^*g\>=0$ and $\<g^*f\>=0$ by the
Cauchy-Schwarz inequality, hence 
$\<(f+g)^*(f+g)\>=\<f^*f\>+\<g^*g\>=0$, so that $f+h\in V$.

If $g\in V$ and $f$ is bounded then 
\[
(fg)^*(fg)=g^*f^*fg\leq g^*\|f\|^2g=\|f\|^2g^*g
\]
implies $\<(fg)^*(fg)\>\leq\|f\|^2\<g^*g\>=0$, so that $fg\in V$.

And if $f\in V$ is Hermitian then $\<f^2\>=\<f^*f\>=0$, and, again
by Cauchy-Schwarz, $\<f^4\>\leq\<f^6\>\<f^2\>=0$, so that $f^2\in V$.
\epf

\section{Probability} 
\label{probability}

\hfill\parbox[t]{8.8cm}{\footnotesize

{\em Enough, if we adduce probabilities as likely as 
any others; for we must remember that I who am the speaker, and you 
who are the judges, are only mortal men, and we ought to accept the 
tale which is probable and enquire no further.}

Plato, ca. 367 B.C. \cite{Pla} 
}\nopagebreak

\bigskip
The interpretation of probability has been surrounded by philosophical
puzz\-les for a long time. {\sc Fine} \cite{Fin} is probably still the 
best discussion of the problems involved; {\sc Hacking} \cite{Hac}
gives a good account of its early history. (See also 
{\sc Home \& Whitaker} \cite{HomW}.) {\sc Sklar} \cite{Skl}
has an in depth discussion of the specific problems relating to 
statistical mechanics. 

Our definition generalizes
the classical intuition of probabilities as weights in a weighted 
average and is modeled after the formula for finite 
probability theory in Example \ref{ex.5.3}(i).
In the special case when a well-defined counting process may be 
associated with the statement whose probability is assessed, our 
exposition supports the conclusion of {\sc Drieschner} \cite[p.73]{Dri},
{\em ``probability is predicted relative frequency''}
(German original: ``Wahrscheinlichkeit ist vorausgesagte relative 
H\"au\-fig\-keit''). More specifically, we assert that, 
{\em for counting events, the probability carries the information of 
expected relative frequency} (see Theorem \ref{t1.5}(iii) below). 

To make this precise we need a precise concept of independent events 
that may be counted. To motivate our definition, assume that we look at
times $t_1,\dots,t_N$ for the presence of an event of the sort we want
to count. We introduce quantities $e_l$ whose value is the amount
added to the counter at time $t_l$. For correct counting, we need
$e_l\approx 1$ if an event happened at time $t_l$, and $e_l\approx 0$
otherwise; thus $e_l$ should have the two possible values $0$ and $1$ 
only. Since these numbers are precisely the Hermitian idempotents
among the constant quantities, this suggests to identify events with 
general Hermitian idempotent quantities. 

In addition, it will be useful to have the more general concept of 
`effects' (cf. {\sc Busch} et al. \cite{BusGL,BusLM}, 
{\sc Davies} \cite{Dav}, {\sc Peres} \cite{Per3})
for more fuzzy, event-like things. 

\begin{dfn}~\nopagebreak

(i) A quantity $e \in \Ez$ satisfying $0\leq e\leq 1$ is called an 
{\bf effect}. The number 
$\<e\>$ is called the {\bf probability} of the effect $e$.
Two effects $e,e'$ are called {\bf independent} in an ensemble 
$\< \cdot \>$ if they commute and satisfy
\[
  \< ee' \> = \< e \> \< e' \>.
\]

(ii) A quantity $e \in \Ez$ satisfying $e^2 = e = e^*$ is called an 
{\bf event}. Two events $e,e'$ are called {\bf disjoint} if 
$ee'=e'e=0$.

(iii) An {\bf alternative} is a family $e_l$ ($l\in L$) of effects such
that 
\[
\sum_{l\in L} e_l \leq 1.
\]
\end{dfn}

\begin{prop}~

(i) Every event is an effect.

(ii) The probability of an effect $e$ satisfies $0\leq\<e\>\leq 1$.

(iii) The set of all effects is convex and closed in the uniform 
topology.

(iv) Any two events in an alternative are disjoint.

\end{prop}

\bepf
(i) holds since $0\leq e^*e=e^2=e$ and 
$0\leq(1-e)^*(1-e)=1-2e+e^2=1-e$. 

(ii) and (iii) follow easily from Proposition \ref{p5.2}.

(iv) If $e_k,e_l$ are events in an alternative then $e_k\leq 1-e_l$
and
\[
(e_ke_l)^*(e_ke_l)=e_l^*e_k^*e_ke_l=e_l^*e_k^2e_l=e_l^*e_ke_l
\leq e_l^*(1-e_l)e_l=0.
\]
Hence $e_ke_l=0$ and $e_le_k=e_l^*e_k^*=(e_ke_l)^*=0$.
\epf

Note that we have a well-defined notion of probability though the
concept of a probability distribution is absent. It is neither needed 
nor definable in general. Nevertheless, the theory contains classical
probability theory as a special case.

\begin{expls}\label{ex1.4}~

(i) {\bf Classical probability theory.} 
In classical probability theory, quantities are usually called 
{\bf random variables}; they belong to the Q-algebra $B(\Omega)$ of 
measurable complex-valued functions on a measurable set $\Omega$. 

The characteristic function $e = \CHI_M$ of any measurable subset $M$ 
of $\Omega$ (with $\CHI_M (\omega ) =1$ if $\omega  \in M$, 
$\CHI_M (\omega )=0$ otherwise) is an event.
A family of characteristic functions $\CHI_{M_l}$ form an alternative 
iff their supports $M_l$ are pairwise disjoint (apart from a set of 
measure zero). 

Effects are the measurable functions $e$ with values in $[0,1]$; they 
can be considered as `characteristic functions' of a fuzzy set where 
$\omega \in\Omega$ has $e(\omega )$ as degree of membership (see, e.g., 
{\sc Zimmermann} \cite{Zim}).

For many applications, the algebra $B(\Omega)$ is too big, and 
suitable subalgebras $\Ez$ are selected on which the relevant 
ensembles can be defined as integrals with respect to suitable positive
measures.

(ii)  {\bf Quantum probability theory.} 
In the algebra of bounded linear operators on a Hilbert space 
$\Hz$, every unit vector $\phi \in \Hz$ gives rise to an 
event $e_\phi = \phi\phi ^*$. We shall call such events 
{\bf irreducible events}. 
A family of irreducible events $e_{\phi_l}$ form an alternative 
iff the $\phi_l$ are pairwise orthogonal. The probability of an
irreducible event $e_\phi$ in an ensemble corresponding to the
unit vector $\psi$ is 
\lbeq{e.sqprob}
\<e_\phi\>=\psi^*e_\phi\psi=\psi^*\phi\phi^*\psi=|\phi^*\psi|^2.
\eeq
This is the well-known {\bf squared probability amplitude} formula,
traditionally interpreted as the probability that after preparing a 
pure ensemble in the pure `state' $\psi$, an `ideal measurement' 
causes a `state reduction' to the new pure `state' $\phi$. 

In contrast, our interpretation of $|\phi^*\psi|^2$ is completely 
within the formal framework of the theory and completely independent 
of the measurement process.

Further, reducible, quantum events are orthogonal projectors to 
subspaces. The effects are the Hermitian operators $e$ with spectrum 
in $[0,1]$.
\end{expls}

\begin{thm}\label{t1.5}~

(i) For any effect $e$, its {\bf negation} $\neg e = 1 - e$ is an
effect with probability 
\[
  \< \neg e \> = 1 - \< e \>;
\]
it is an event if $e$ is an event.

(ii) For commuting effects $e, e'$, the quantities
\[
  e \wedge e' = ee' ~~~(e \mbox{ \bf and } e'),
\]
\[
  e \vee e' = e + e' - ee' ~~~(e \mbox{ \bf or } e')
\]
are effects whose probabilities satisfy
\[
  \< e \wedge e' \> + \< e \vee e' \> = \< e \> + \< e' \>;
\]
they are events if $e, e'$ are events. Moreover,
\[
\<e\wedge e'\> = \<e\>\<e'\>~~~\mbox{for independent effects } e, e'.
\]

(iii) For a family of effects $e_l$ $(l=1, \ldots , N)$ with
constant probability $\< e_l \> = p$, the {\bf relative frequency}
\[
  q := \frac{1}{N} \D \sum ^N _{l=1} e_l
\]
satisfies
\[
  \< q \> =p.
\]
(iv) For a family of independent events of probability $p$,
the uncertainty 
\[
\sigma (q) = \sqrt{ \frac{ p(1-p)}{N}}
\]
of the relative frequency becomes arbitrarily small as $N$ becomes 
sufficiently large {\bf (weak law of large numbers)}.
\end{thm}

\bepf
(i) $\neg e$ is an effect since $0\leq 1-e\leq1$, and its probability
is $\< \neg e \> = \< 1-e \> = 1 - \< e \>$.
If $e$ is an event then clearly $\neg e$ is Hermitian, and 
$(\neg e)^2=(1-e)^2=1-2e+e^2=1-e=\neg e$. Hence $\neg e$ is an event.

(ii) Since $e$ and $e'$ commute, $e \wedge e'=ee'=e^2e'=ee'e$.
Since $ee'e\geq0$ and $ee'e\leq ee=e\leq1$, we see that $e \wedge e'$
is an effect. Therefore, 
$e \vee e'=e+e'-ee'=1-(1-e)(1-e')=\neg (\neg e\wedge \neg e')$ 
is also an effect. The assertions about expectations are immediate.
If $e,e'$ are events then $(ee')^* =e'^* e^* =e'e=ee'$, hence 
$ee'$ is Hermitian; and it is idempotent since 
$(ee')^2=ee'ee'=e^2e'^2=ee'$. Therefore $e \wedge e'=ee'$ is an
event, and $e \vee e'=\neg (\neg e\wedge \neg e')$ is an event, too.

(iii) This is immediate by taking the expectation of $q$.

(iv) This follows from Theorem \ref{t.weaklaw} since
$\<e_k^2\>=\<e_k\>=p$ and 
\[
\sigma(e_k)^2= \< (e_k-p)^2\> =\< e_k^2\> -2p\< e_k\> +p^2 
=p-2p^2+p^2=p(1-p).
\]
\epf

We remark in passing that, with the operations $\wedge,\vee,\neg$, 
the set of events in any {\em commutative} subalgebra of $\Ez$ 
forms a Boolean algebra; see {\sc Stone} \cite{Sto}. Traditional
quantum logic (see, e.g., {\sc Birkhoff \& von Neumann} \cite{BirN}, 
{\sc Pi\-tow\-sky} \cite{Pit}, {\sc Svozil} \cite{Svo}) discusses the 
extent to which this can be generalized to the noncommutative case. 
We shall make no use of quantum logic; the only logic used is 
classical logic, applied to well-defined assertions about quantities.
However, certain facets of quantum logic related to so-called 
`hidden variables' are discussed from a different point of view in the 
next section.

The set of effects in a commutative subalgebra is {\em not} a 
Boolean algebra. Indeed, $e \wedge e \neq e$ for effects $e$ that are 
not events. In fuzzy set terms, if $e$ codes the answer to the question 
`(to which degree) is statement $S$ true?' then $e \wedge e$ codes the 
answer to the question `(to which degree) is statement $S$ really 
true?', indicating the application of more stringent criteria for 
truth. See {\sc Neumaier} \cite{Neu.surprise} for a more rigorous
discussion of this aspect.

For noncommuting effects, `and' and `or' ar undefined. One might 
think of $\shalf(ee'+e'e)$ as a natural definition for $e \wedge e'$;
however, this expression need not be an effect (not even when both 
$e$, $e'$ are events), as the following simple example shows:
\[
e=\left(\bary{cc}1&0\\0&0 \end{array}\right),~~
e'=\half\left(\bary{cc}1&1\\1&1 \end{array}\right),~~~~
\half(ee'+e'e)=\frac{1}{4}\left(\bary{cc}2&1\\1&0 \end{array}\right).
\]

\bigskip
\section{Nonlocality} 
\label{nonlocality}

\hfill\parbox[t]{8.8cm}{\footnotesize

{\em As the heavens are higher than the earth, so are my ways higher 
than your ways and my thoughts than your thoughts.}

The {\sc LORD}, according to Isaiah, ca. 540 B.C. \cite{Isa} 

\bigskip
{\em Before they call I will answer; while they are still speaking 
I will hear.}

The {\sc LORD}, according to Isaiah, ca. 540 B.C. \cite{Isa2} 
}\nopagebreak

\bigskip
A famous feature of quantum physics is its intrinsic nonlocality.

\begin{expl}\label{spinpair}
In $\Cz^{4 \times 4}$, the four matrices $f_j$ defined by
\[
f_1x=\left(\begin{array}{r}x_3\\x_4\\x_1\\x_2\end{array}\right),~
f_2x=\left(\begin{array}{r}x_2\\x_1\\x_4\\x_3\end{array}\right),~
f_3x=\left(\begin{array}{r}x_1\\x_2\\-x_3\\-x_4\end{array}\right),~
f_4x=\left(\begin{array}{r}x_1\\-x_2\\x_3\\-x_4\end{array}\right)
\]
satisfy 
\lbeq{e.bell1}
f_k^2\leq 1~~~\mbox{for } k=1,2,3,4,
\eeq
and $f_j$ and $f_k$ commute for odd $j-k$. It is easily checked that
in the pure ensemble defined by the vector 
\[
\psi=
\left(\begin{array}{r}
\alpha_1\\-\alpha_2\\ \alpha_2\\ \alpha_1
\end{array}\right),
~~~\alpha_{1,2}=\sqrt{\frac{2\pm\sqrt{2}}{8}},
\]
we have
\lbeq{e.bell1a}
\<f_1f_2\>=\<f_3f_2\>=\<f_3f_4\>=-\<f_1f_4\>=\half\sqrt{2}. 
\eeq
Since $\<f_k\>=0$ for all $k$, this implies that $f_j$ and $f_k$ are 
correlated for odd $j-k$. On identifying 
\[
\left(\begin{array}{l}x_1\\x_2\\x_3\\x_4\end{array}\right)
=\left(\begin{array}{l}x_1~~x_2\\x_3~~x_4\end{array}\right)
\]
and defining the tensor product action $u\otimes v: x \mapsto uxv^T$, 
the matrices $f_j$ can be written in terms of the Pauli spin matrices 
\gzit{e.pauli} as
\[
f_1 = \sigma_1 \otimes 1,~~f_2 = 1 \otimes \sigma_1,~~
f_3 = \sigma_3 \otimes 1,~~f_4 = 1 \otimes \sigma_3.
\]
If we interpret the two terms in a tensor product as quantities related 
to two spatially separated Fermion particles $A$ and $B$, we conclude
from \gzit{e.bell1a}
that there are pure ensembles in which the components of the spin 
vectors of two fermion particles are necessarily correlated, 
no matter how far apart the two particles are placed, and no matter 
what was their past. 
Such {\em nonlocal correlations} of certain quantum ensembles are
an enigma of the microscopic world that, being experimentally 
confirmed, cannot be removed by any interpretation of quantum
mechanics.
\end{expl}

The nonlocal properties of quantum mechanics are usually
expressed by so-called {\bf Bell inequalities} 
(cf. {\sc Bell} \cite{Bel}, {\sc Clauser \& Shimony} \cite{ClaS}). 
The formulation given here depends on the most orthodox part of 
quantum mechanics only; unlike in most expositions, it neither refers
to hidden variables nor involves counterfactual reasoning.

\begin{thm}\label{t.bell}
Let $f_k$ ($k=1,2,3,4$) be Hermitian quantities satisfying 
\gzit{e.bell1}.

(i) {\rm (cf. {\sc Cirel'son} \cite{Cir})} 
For every ensemble,
\lbeq{e.bellq}
|\<f_1f_2\>+\<f_3f_2\>+\<f_3f_4\>-\<f_1f_4\>|\leq 2\sqrt{2}.
\eeq

(ii) {\rm (cf. {\sc Clauser} et al. \cite{ClaHS})}
If, for odd $j-k$, the quantities $f_j$ and $f_k$ commute and are 
uncorrelated then we have the stronger inequality
\lbeq{e.bellc}
|\<f_1f_2\>+\<f_3f_2\>+\<f_3f_4\>-\<f_1f_4\>|\leq 2.
\eeq

\end{thm}

\bepf
(i) Write $\gamma$ for the left hand side of \gzit{e.bellq}. Using
the Cauchy-Schwarz inequality and the easily verified inequality 
\[
\sqrt{\alpha}+\sqrt{\beta}\leq\sqrt{2(\alpha+\beta)}~~~
\mbox{for all } \alpha, \beta \geq 0, 
\]
we find
\[
\begin{array}{lll}
\gamma&=&|\<f_1(f_2-f_4)\>+\<f_3(f_2+f_4)\>| \\
&\leq&\sqrt{\<f_1^2\>\<(f_2-f_4)^2\>}+\sqrt{\<f_3^2\>\<(f_2+f_4)^2\>} \\
&\leq&\sqrt{\<(f_2-f_4)^2\>}+\sqrt{\<(f_2+f_4)^2\>} \\
&\leq&\sqrt{2(\<(f_2-f_4)^2\>+\<(f_2+f_4)^2\>)} 
       =\sqrt{4\<f_2^2+f_4^2\>}=2\sqrt{2}. \\
\end{array}
\]

(ii) By Proposition \ref{p5.2}(ii), $v_k:=\<f_k\>$ satisfies
$|v_k|\leq 1$. If $f_j$ and $f_k$ commute for 
odd $j-k$ then Proposition \ref{p5.2}(iv) implies $\<f_jf_k\>=v_jv_k$ 
for odd $j-k$. Hence
\[
\begin{array}{lll}
\gamma&=&|v_1v_2+v_3v_2+v_3v_4-v_1v_4| =|v_1(v_2-v_4)+v_3(v_2+v_4)| \\
&\leq&|v_1|~|v_2-v_4|+|v_3|~|v_2+v_4| \leq |v_2-v_4|+|v_2+v_4| \\
&=&2\max(|v_2|+|v_4|)\leq 2. \\
\end{array}
\]
\epf

For instance, in the above example, \gzit{e.bell1a} implies that 
\gzit{e.bellq} holds with equality but \gzit{e.bellc} is violated.
Indeed, the assumption of (ii) is not satisfied.

The significance of the theorem stems from the fact that it implies 
that it is impossible to prepare a classical ensemble for which 
the $f_i$ have the same correlations as in Example \ref{spinpair}, 
thus excluding the existence of local hidden variable theories.

See {\sc Bell} \cite{Bel} for the original Bell inequality,
{\sc Pitowsky} \cite{Pit} for a treatise on Bell inequalities,
and {\sc Aspect} \cite{Asp}, {\sc Clauser \& Shimony} \cite{ClaS}, 
{\sc Tittel} et al. \cite{TitBG} for experiments
verifying the violation of \gzit{e.bellc}.

\section{Experiments} \label{ideal}

\hfill\parbox[t]{9.9cm}{\footnotesize

{\em A phenomenon is not yet a phenomenon until it has been brought 
to a close by an irreversible act of amplification [...] 
What answer we get depends on the question we put, the experiment 
we arrange, the registering device we choose.}

John Archibald Wheeler, 1981 \cite{Whe}
}\nopagebreak

\bigskip

The literature on the foundations of physics is full of discussions of
the measurement process ({\sc Wheeler \& Zurek} \cite{WheZ}), 
usually in a heavily idealized fashion
(which might well be responsible for the resulting paradoxa).
Measurements such as that of the mass of top quarks or of neutrinos
have a complexity that in no way is covered by the traditional 
foundational measurement discussions. To a lesser degree, this is also
true of most other measurements realized in modern physics. 

Indeed, the values of quantities of interest are usually obtained by a 
combination of observations and calculations. For science, it is of 
utmost importance to have well-defined protocols that specify how to 
arrive at valid observations. Such standardized protocols guarantee 
that the observations have a high degree of reproducibility and hence 
are objective. 

On the other hand, these protocols require a level of 
description not appropriate for the foundations of a discipline.
(E.g., we read a number from a meter and claim having measured something
only indirectly related to the meter through theory far away from the
foundations.)

We shall therefore formalize the notion of an experiment by taking
into account only their most obvious aspect, and consider experiments 
to be assignments of complex numbers $v(f)$ to certain quantities $f$. 
This abstracts the results that can be calculated from an experiment
without entering the need to discuss details of how such experiments
can be performed. In particular, the thorny issue of how the process
of performing an experiment affects the observations can be excluded
from the foundations.

Our rudimentary but precise notion of experiment is sufficient to 
discuss consistency conditions that describe how `good' experiments 
should relate to a physical theory, thus separating the basics from 
the complications due to real experiments. 

Since not all experiments allow one to assign values to all 
quantities, we need a symbol `?' that indicates an unspecified 
(and perhaps undefined) value. Operations involving ? give ? as a 
result, with exception of the rule
\[
0?=?0=0.
\]
Apart from this, we demand minimal requirements shared by all 
reasonable assignments in an experiment. For `good' experiments,
additional constraints should be imposed -- which ones are most
meaningful will be analyzed in the following. In particular, we look
at the constraint imposed by `sharpness'.

\begin{dfn}~

(i) A {\bf experiment} is a mapping $v:\Ez \to \Cz \cup \{ ? \}$
such that

(S1) $v(\alpha+\beta f)=\alpha+\beta  v(f)$~~~ if $\alpha, \beta\in\Cz$,

(S2) $v (f)\in\Rz\cup\{?\}$ ~~~if $f$ is Hermitian;

it is called {\bf complete} if $v (f)\in\Cz$ for all $f\in\Ez$.
$v(f)$ is called the {\bf reference value} of $f$ in the experiment $v$.
\[
\Ez_v:= \{ f\in \Ez \mid v (f)\in \Cz \}
\]
denotes the set of quantities with definite values in experiment $v$.

(ii) A set $E$ of Hermitian quantities is called {\bf sharp} in 
experiment $v$ if, for $f,g \in E$ and $\lambda\in\Rz$,,

(SQ0) $\Rz \subseteq E,~~~v(f)\in\Rz$,

(SQ1) $f^2 \in E,~~~ v(f^2) = v(f)^2$,
 
(SQ2) $f^{-1}\in E$,~~~$v(f^{-1})=v(f)^{-1}$ ~~~if $f$ is invertible,

(SQ3) $f\pm g \in E,~ v(f+\lambda g) = v(f) + \lambda v(g)$ 
~~~if $f,g$ commute.

A quantity $f$ is called {\bf sharp} in $v$ if $\re f$ and $\im f$ 
commute and belong to some set that is sharp in experiment $v$.

\end{dfn}

Thus, sharp quantities behave with respect to their reference values
precisely as numbers would do (hence the name). In particular, sharp 
quantities are normal by Corollary \ref{c1.3}.

\begin{expls}\label{ex.classquant}
According to tradition, the best an experiment can do is to detect the
location of a classical system in phase space, or the wave function 
of a quantum system. These ideal experiments are describes in the
present setting as follows.

(i) {\bf Classical mechanics.} 
Classical few-particleparticle mechanics with $N$ degrees of freedom
is described by a {\em phase space} $\Omega_{cl}$, 
the direct product of $\Rz^N\times\Rz^N$ and a compact 
manifold describing internal particle degrees of freedom. 
$\Ez$ is a subalgebra of the algebra $B (\Omega_{cl})$ of
Borel measurable functions on {\bf phase space} $\Omega_{cl}$.

A {\bf classical point experiment} is determined by a phase space point
$\omega\in\Omega_{cl}$ and the recipe
\[
  v_\omega (f) := 
   \left\{ \bary{ll}
     f(\omega) & \mbox{if  $f$ is continuous at $\omega$},\\
     ? & \mbox{otherwise.}
   \eary \right.
\]
In a classical point experiment $v$, all $f\in\Ez_v$ are sharp 
(and normal). 

In classical probability theory, a point experiment is 
usually referred to as a {\bf realization}.

(ii) {\bf Nonrelativistic quantum mechanics.}
Nonrelativistic quantum mechanics of $N$ particles is described by a 
Hilbert space $\Hz=L^2(\Omega_{qu})$, where $\Omega_{qu}$ is the 
direct product of $\Rz^N$ and a finite set that takes care of spin, 
color, and similar indices. $\Ez$ is the algebra of 
bounded linear operators on $\Hz$. 
(If unbounded operators are considered, $\Ez$ is instead an algebra of 
linear operators in the corresponding Schwartz space, but for this 
example, we do not want to go into technical details.) 

The Copenhagen interpretation is the most prominent, and at the same 
time the most restrictive, interpretation of quantum mechanics. 
It assigns definite values only to quantities in an eigenstate.
A {\bf Copenhagen experiment} is determined by a wave function
$\psi \in \Hz\setminus \{0\}$ and the recipe
\[
  v _\psi (f) := 
   \left\{ \bary{ll}
     \lambda & \mbox{if } f \psi = \lambda \psi, \\
     ? & \mbox{otherwise.}
   \eary \right.
\]
In a Copenhagen experiment $v$, all normal $f\in\Ez_v$ 
(defined in (S2)) are sharp.
\end{expls}

While well-defined value assignments model the repeatability
of an experiment and are indispensable in any objective theory, 
sharpness is a matter not of objectivity but one of `point-like' 
specificity of the assignments. Thus the extent to which sharpness
can be consistently assumed reflects a property of the real world.
Indeed, sharpness is the traditional characteristics of a classical 
world with a commutative algebras of quantities. 

We now investigate the properties 
of sharp quantities in general experiments.
It will turn out that total sharpness is incompatible with the
existence of pairs of spins: no experiment -- in the very general
setting defined here -- can give sharp values to all Hermitian 
quantities. Hence total sharpness contradicts our knowledge of the 
world, another expression of the nonlocal nature of reality.

Our first observation is that numbers are their own reference values, 
and that sharp events are dichotomic -- their only possible reference 
values are $0$ and $1$.

\begin{prop}~

(i)~ $v(\alpha)=\alpha$~ if $\alpha \in \Cz$.

(ii) If $e$ is a sharp event then $v(e)\in\{0,1\}$.
\end{prop}

\bepf
(i) is the case $\beta=0$ of (S1), and (ii) holds since
in this case, (SQ1) implies $v(e)=v(e^2)=v(e)^2$.
\epf

\begin{prop}
If the set $E$ is sharp in the experiment $v$ then
\lbeq{e.s1}
fg\in E,~~v(fg) = v(f)v(g)~~~\mbox{if $f,g \in E$ commute},
\eeq
\lbeq{e.s2}
\alpha+\beta f \in E,~~v(\alpha+\beta f)=\alpha+\beta v(f)
~~~\mbox{if } f \in E,\alpha,\beta\in\Rz.
\eeq
\end{prop}

\bepf
If $f,g \in E$ commute then $f\pm g \in E$ by (SQ3). By (SQ1),
$(f\pm g)^2\in E$ and $v((f\pm g)^2)=v(f\pm g)^2$. By (SQ3),
$fg=((f+g)^2-(f-g)^2)/4$ belongs to $E$ and satisfies
\[
\begin{array}{lll}
4v(fg)&=&v((f+g)^2)-v((f-g)^2)=v(f+g)^2-v(f-g)^2\\
&=&(v(f)+v(g))^2-(v(f)-v(g))^2=4v(f)v(g).
\end{array}
\]
Thus \gzit{e.s1} holds, and \gzit{e.s2} 
follows from \gzit{e.s1}, (SQ0) and (SQ3). 
\epf

\bigskip
One of the nontrivial traditional {\em postulates} of quantum mechanics,
that the possible values a sharp quantity $f$ may take are the 
elements of the spectrum $\spec f$ of $f$, is a {\em consequence} of 
our axioms. 

\begin{thm}\label{t.spec}
If a Hermitian quantity $f$ is sharp in the experiment $v$, and 
$v(f)=\lambda$ then: 

(i) $\lambda-f$ is not invertible. 

(ii) If there is a polynomial $\pi(x)$ such that $\pi(f)=0$
then $\lambda$ satisfies $\pi(\lambda)=0$. In particular,
if $f$ is a sharp event then $v(f)\in \{0,1\}$.

(iii) If $\Ez$ is finite-dimensional then there is a quantity $g\neq 0$
such that $fg=\lambda g$, i.e., $\lambda$ is an eigenvalue of $f$.

\end{thm}

\bepf
Note that $\lambda$ is real by (SQ0).

(i) If $g:=(\lambda-f)^{-1}$ exists then by \gzit{e.s2} and (SQ2),
$\lambda-f,g\in E$ and
\[
v(\lambda-f)v(g)=v((\lambda-f)g)=v(1)=1,
\]
contradicting $v(\lambda-f)=\lambda-v(f)=0$.

(ii) By polynomial division we can find a polynomial $\pi_1(x)$ such 
that $\pi(x)=\pi(\lambda)+(x-\lambda)\pi_1(x)$. If $\pi(\lambda)\neq 0$,
$g:=-\pi_1(f)/\pi(\lambda)$ satisfies 
\[
(\lambda-f)g=(f-\lambda)\pi_1(f)/\pi(\lambda)
=(\pi(\lambda)-\pi(f))/\pi(\lambda)=1,
\]
hence $\lambda-f$ is invertible with inverse $g$, contradiction. 
Hence $\pi(\lambda)=0$. In particular, this applies to an event with 
$\pi(x)=x^2-x$; hence its possible reference values are zeros of 
$\pi(x)$, i.e., either $0$ or $1$.

(iii) The powers $f^k$ ($k=0,\dots, \dim \Ez$) must be linearly 
dependent; hence there is a polynomial $\pi(x)$ such that $\pi(f)=0$.
If this is chosen of minimal degree then $g:=\pi_1(f)$ is nonzero
since its degree is too small. Since
$0=\pi(\lambda)=\pi(f)+(f-\lambda)\pi_1(f)=(f-\lambda)g$, we have
$fg=\lambda g$.
\epf

When $\Ez$ is a $C^*$-algebra, the spectrum of $f\in \Ez$ is defined 
as the set of complex numbers $\lambda$ such that $\lambda-f$ has no 
inverse (see, e.g., \cite{Ric}). Thus in this case, part (i) of the 
theorem implies that all numerical values a sharp quantity $f$ can 
take belong to the spectrum of $f$. This covers both the case of 
classical mechanics and that of nonrelativistic quantum mechanics.

\bigskip
Sharp quantities always satisfy a {\bf Bell inequality} analogous to 
inequality \gzit{e.bellc} for uncorrelated quantities:

\begin{thm}\label{t.bells}
Let $v$ be an experiment with a sharp set of quantities containing four
Hermitian quantities $f_j$ ($j=1,2,3,4$) satisfying $f_j^2=1$
and $[f_j,f_k]=0$ for odd $j-k$. Then 
\lbeq{e.bellcs}
|v(f_1f_2)+v(f_2f_3)+v(f_3f_4)-v(f_1f_4)|\leq 2.
\eeq

\end{thm}

\bepf
Let $v_k:=v(f_k)$. Then (SQ2) implies $v_k^2=v(f_k^2)=v(1)=1$, and
since equation \gzit{e.s1} implies  $v(f_jf_k)=v_jv_k$ for odd
$j-k$, we find
\[
\begin{array}{lll}
\gamma&=&|v_1v_2+v_2v_3+v_3v_4-v_1v_4| \\
&=&|v_1(v_2-v_4)+v_3(v_2+v_4)| \\
&\leq&|v_1|~|v_2-v_4|+|v_3|~|v_2+v_4| \\
&\leq&|v_2-v_4|+|v_2+v_4| \\
&=&2\max(|v_2|+|v_4|)\leq 2. \\
\end{array}
\]
\epf

\section{Which assumptions?}

\hfill\parbox[t]{11.4cm}{\footnotesize

{\em That so much follows from such apparently innocent assumptions 
leads us to question their innocence.}

John Bell, 1966 \cite{Bel3} 

\bigskip
{\em For example, nobody doubts that at any given time the center of 
mass of the Moon has a definite position, even in the absence of any 
real or potential observer.}

Albert Einstein, 1953 \cite{Ein2} 
}\nopagebreak

\bigskip

In this section we discuss the question: Assuming there is a
consistent, objective physical reality behind quantum physics 
which can be described by precise mathematics, what form can it take?

Taking 'physical reality' in our mathematical model as synonymous with
'being observable in an experiment', the question becomes one of
finding natural consistency conditions for experiments which
assign to {\em all} quantities reference values which could qualify 
as the objective description of physical reality -- the `state' of the system, 
the `beables' of {\sc Bell} \cite{Bel2}.

The most popular conditions posed in the literature are equivalent to 
(or stronger than) sharpness. But, in general, one cannot hope that 
{\em every} Hermitian quantity is sharp. Indeed, it was shown by 
{\sc Kochen \& Specker} \cite{KocS} that there is a finite set of 
events in $\Cz^{3 \times 3}$ (and hence in any Hilbert space of 
dimension $>2$) for which {\em any} assignment of reference values 
leads to a contradiction with the sharpness conditions. 
We give a slightly less general result that is much easier to prove.

\begin{thm}\label{t.nocons}
{\rm (cf. {\sc Mermin} \cite{Mer}, {\sc Peres} \cite{Per2})}
\\\nopagebreak
There is no experiment with a sharp set of quantities containing four
Hermitian quantities $f_j$ ($j=1,2,3,4$) satisfying $f_j^2=1$ and
\lbeq{e.nocons}
f_jf_k=\left\{ \begin{array}{rl}
-f_kf_j &\mbox{if } j-k=\pm 2,\\
f_kf_j &\mbox{otherwise}.
\end{array}\right.
\eeq
\end{thm}

\bepf
Let $E$ be a set containing the $f_j$. If $E$ is sharp in the 
experiment $v$
then $v_j=v(f_j)$ is a number, and $v_j^2=v(f_j^2)=v(1)=1$ implies 
$v_j \in\{-1,1\}$. In particular,
$v_0:=v_1 v_2 v_3 v_4\in\{-1,1\}$.
By \gzit{e.s1}, $v(f_j f_k)=v_j v_k$ if $j,k\neq \pm 2$. 
Since $f_1f_2$ and $f_3f_4$ commute,
$v(f_1f_2f_3f_4)=v(f_1f_2) v(f_3f_4)=v_1v_2v_3 v_4
=v_0$, and since $f_1f_4$ and $f_2f_3$ commute,
$v(f_1f_4f_2f_3)=v(f_1f_4)v(f_2f_3) v_1v_4v_2 v_3
=v_0$. Since $f_1f_4f_2f_3=-f_1f_2f_3f_4$, this gives 
$v_0=-v_0$, hence the contradiction $v_0=0$.
\epf

\begin{expl}\label{ex.peres}
The $4 \times 4$-matrices $f_j$ defined in Example \ref{spinpair} 
satisfy the relations required in Theorem \ref{t.nocons}. 
As a consequence, there cannot be an experiment 
in which all components of the spin vectors of two Fermions are sharp.
\end{expl}

This implies that the 
sharpness assumption in Theorem \ref{t.bells} and in other 
Bell-type inequalities for local hidden variable theories 
(see, e.g., the treatise by {\sc Pitowsky} \cite{Pit}) fails not only 
in special entangled ensembles such as that exhibited in 
Example \ref{spinpair} but must {\em fail independent of any special 
preparation.}

A similar interpretation can be given for a number of other arguments 
against so-called local hidden variable theories, which assume that 
{\em all} Hermitian quantities are sharp. 
(See {\sc Bernstein} \cite{Ber}, {\sc Eberhard} \cite{Ebe}, 
{\sc Greenberger} et al. \cite{GreHS,GreHZ}, 
{\sc Hardy} \cite{Har,Har2}, {\sc Mermin} \cite{Mer,Mer2}, 
{\sc Peres} \cite{Per,Per2}, {\sc Vaidman} \cite{Vai}). 
For a treatment in terms of quantum logic, see {\sc Svozil} \cite{Svo}.

While the above results show that one cannot hope to find quantum 
experiments in which all Hermitian quantities are sharp, results of
{\sc Clifton \& Kent} \cite{CliK} imply that one can achieve sharpness 
in $\Ez=\Cz^{n\times n}$ at least for a dense subset of Hermitian 
quantities.

\bigskip
Since, as we have seen, experiments in which all Hermitian quantities 
are sharp are impossible, we need to discuss the relevance of the 
sharpness assumption for reference values that characterize 
experiments.

The chief culprit among the sharpness assumptions seems to be the
squaring rule (SQ1) from which the product rule \gzit{e.s1} was
derived. Indeed, the squaring rule (and hence the product rule) already 
fails in a simpler, classical situation, namely when considering 
weak limits of highly oscillating functions, 
For example, consider the family of functions $f_k$ defined on $[0,1]$
by $f_k(x)=\alpha$ if $\lfloor kx \rfloor$ is even and $f_k(x)=\beta$ 
if $\lfloor kx \rfloor$ is odd. Trivial integration shows that  
the weak-$^*$ limits are $\lim f_k=\shalf(\alpha+\beta)$ and 
$\lim f_k^2=\shalf(\alpha^2+\beta^2)$, and these do not satisfy the 
expected relation $\lim f_k^2= (\lim f_k)^2$. Such weak limits of 
highly oscillating functions lead to the concept of a 
{\em Young measure}, which is of relevance in the calculus of 
variation of nonconvex functionals and in the physics of metal 
microstructure. See, e.g., {\sc Roubicek} \cite{Rou}.

More insight from the classical regime comes from realizing that 
reference values are a microscopic analogue of similar macroscopic 
constructions. 

For example, the center of mass, the mass-weighted average of the 
positions of the constituent particles, serves in classical mechanics 
as a convenient reference position of an extended object. It defines a 
point in space with a precise and objective physical meaning. 
The object is near this reference position, within an uncertainty 
given by the diameter of the object. Similarly, a macroscopic object 
has a well defined reference velocity, the mass-weighted average of 
the velocities of the constituent particles.

Thus, if we define an algebra $\Ez$ of `intensive' macroscopic 
mechanical quantities, given by all (mass-independent and sufficiently 
nice) functions of time $t$, position $q(t)$, velocity $\dot q(t)$ 
and acceleration $\ddot q(t)$, the natural reference value 
$v_{mac}(f)$ for a quantity $f$ is the mass-weighted average of the 
$f$-values of the constituent particles (labeled by superscripts $a$),
\[
v_{mac}(f)= 
\sum_a m^a f(t, q^a(t),\dot q^a(t), \ddot q^a(t))\Big/\sum_a m^a.
\]
This reference value behaves correctly under aggregation, if on the 
right hand side the reference values of the aggregates are substituted,
so that it is independent of the details of how the object is split 
into constituents. Moreover, $v=v_{mac}$ has nice properties: 
{\bf unrestricted additivity},

(SL) $v(f+g) = v(f) + v(g)$ ~~~if $f,g \in \Ez$,

and {\bf monotony},

(SM) $f\geq g \implies v(f)\geq v(g)$.

However, neither position nor velocity nor acceleration is a sharp 
quantity with respect to $v_{mac}$ since (SQ1) and (SQ2) fail. 
Note that deviations from the squaring rule make physical sense; 
for example, for an ideal gas in thermodynamic equilibrium,
$v_{mac}(\dot q^2)-v_{mac}(\dot q)^2$ is proportional to the 
temperature of the system. 

From this perspective, and in view of Einstein's quote at the 
beginning of this section, demanding the squaring rule for a 
reference value is unwarranted since it does not even hold in this 
classical situation.

\bigskip
Once the squaring rule (and hence sharpness) is renounced as a 
requirement for definite reference values, the stage is free for 
interpretations that use reference values defined for {\em all} 
quantities, and thus give a satisfying realistic picture of quantum 
mechanics. In place of the lost multiplicative properties we may now 
require unrestricted additivity (SL) without losing interesting 
examples. 

For example, the `local expectation values' in the hidden-variable 
theory of Bohmian mechanics ({\sc Bohm} \cite{Boh}) have this 
property, if the prescription given 
for Hermitian quantities in {\sc Holland} \cite[eq. (3.5.4)]{Hol} is 
extended to general quantities, using the formula
\[
v(f) :=v(\re f) +iv(\im f)
\]
which follows from (SL). Such {\bf Bohmian experiments} have, by 
design, sharp positions at all times. However, they lack 
desirable properties such as monotony (SM), and they display other 
counterintuitive behavior. Moreover, Bohmian mechanics has no natural 
Heisenberg picture, cf. {\sc Holland} \cite[footnote p. 519]{Hol}. 
(The reason is that noncommuting position operators at different times 
are assumed to have sharp values.)

But a much more natural proposal comes from considering the 
statistical foundations of thermodynamics. 

\bigskip
\section{Consistent experiments}\label{consistent}

\hfill\parbox[t]{11.4cm}{\footnotesize

{\em One is almost tempted to assert that 
the usual interpretation in terms of sharp eigenvalues is ``wrong'', 
because it cannot be consistently maintained, while the interpretation 
in terms of expectation values is ``right'', because it can be 
consistently maintained.
}

John Klauder, 1997 \cite{Kla}

\bigskip
{\em This means that the photon must have occupied a volume larger 
than the slit separation. On the other hand, when it fell on the 
photographic plate, the photon must have been localized into the 
tiny volume of the silver embryo.}

Braginsky and Khalili, 1992 \cite{BraK} 
}\nopagebreak

\bigskip
In the derivations of thermodynamics from statistical mechanics, 
it is shown that all extensive 
quantities are expectations in a (grand canonical) ensemble, while 
intensive quantities are parameters in the density determined by the 
extensive quantities and the equation of state. Thus, from the 
macroscopic point of view, ensembles seem to be the right objects 
for defining reference values.
That what we measure {\em reliably} in practice to high accuracy are 
usually also expectations (means, probabilities) points in the same 
direction.

Indeed, each ensemble defines a complete experiment by 
\[
v(f):=\<f\> \forall f \in\Ez,
\]
for which (SL) and (SM) hold. For such experiments one 
even has a meaningful replacement for the multiplicative properties: 
It follows from \gzit{e.prodbound} that 
there is an uncertainty measure 
\lbeq{e.uncmeas}
\Delta f =\sqrt{v(f^2)-v(f)^2}
\eeq
associated with each Hermitian quantity $f$ such that
\lbeq{e.prods}
|v(fg)-v(f)v(g)|\leq\Delta f\Delta g
~~~\mbox{for commuting Hermitian } f,g.
\eeq
Thus the product rule (and in particular the squaring rule) holds in an
approximate form. 

For quantities with small uncertainty $\Delta f$, we have essentially 
classical (nearly sharp) behavior. Im particular,
by the weak law of large numbers (Theorem \ref{t.weaklaw}), averages 
over many uncorrelated commuting quantities of the same kind have small
uncertainty and hence are nearly classical. This holds
for the quantities considered in statistical mechanics, and
{\em explains the emergence of classical properties for macroscopic 
systems}.
Indeed, in statistical mechanics, classical values for observed 
quantities are traditionally defined as expectations, and defining 
objective reference values for all quantities by means of an ensemble 
simply extends this downwards to the quantum domain. 

We therefore call an experiment $v$ {\bf consistent} if 
there is an ensemble $\<\cdot \>$ such that
\[
v(f)\in\{\<f\>,?\} \forall f \in \Ez.
\]
A complete consistent experiment then fully specifies a unique
ensemble and hence the 'state' of the system.

\begin{expls}~

(i) {\bf The ground state of hydrogen.}
The uncertainty $\Delta q$ of the electron position (defined by 
interpreting \gzit{e.uncmeas} for the vector $q$ in place of the 
scalar $f$) in the ground state of hydrogen is 
$\Delta q=\sqrt{3} r_0$ (where $r_0=5.29\cdot 10^{-11}\fct{m}$ is the 
Bohr radius of a hydrogen atom), slightly larger than the reference 
radius $v(r)=\<|q-v(q)|\>=1.5 r_0$. The square of the absolute value 
of the wave function describes the electron as an extended object 
with fuzzy boundaries described by a quickly decaying density, whose
reference position is the common center of mass of nucleus and 
electron.

(ii) {\bf The center of mass of the Moon.}
The Moon has a mass of $m_{\fns{Moon}}=7.35\cdot 10^{22} \fct{kg}$,
Assuming the Moon consists mainly of silicates, we may take the
average mass of an atom to be about 20 times the proton mass
$m_{\fns{p}}=1.67 \cdot 10^{-27} \fct{kg}$. Thus the Moon contains 
about $N=m_{\fns{Moon}}/20m_{\fns{p}}=2.20\cdot 10^{48}$ atoms.
In the rest frame of the Moon, the objective 
uncertainty of an atom position (due to the thermal motion of the 
atoms in the Moon) may be taken to be a small multiple of the Bohr 
radius $r_0$. Assuming that the deviations from the 
reference positions are uncorrelated, we may use \gzit{e.sigN}
to find as uncertainty of the position of the center of mass of the 
Moon a small multiple of $r_0/\sqrt{N}=3.567\cdot 10^{-35}\fct{m}$. 
Thus the center of mass of the Moon has a definite objective 
position, sharp within the measuring accuracy of many generations to 
come. 
\end{expls}

With the assumption that the only experiments consistently realized 
in quantum mechanics are the consistent experiments according to the
above formal definition, the riddles posed by 
the traditional interpretation of the microworld which imagines 
instead pointlike (sharp) properties, are significantly reduced.

Quantum reality with reference values defined by consistent 
experiments is as well-behaved and objective as classical macroscopic 
reality with reference values defined by a mass-weighted average 
over constituent values, and lacks sharpness (in the sense of our 
definition) to the same extent as classical macroscopic reality. 

Consistent experiments provide an elegant solution to the reality 
problem, confirming the insistence of the orthodox Copenhagen 
interpretation on that there is nothing but ensembles, 
while avoiding its elusive reality picture. 
It also conforms to {\sc Ockham}'s razor \cite{Ock, HofMC}, 
{\em frustra fit per plura quod potest fieri per pauciora}, 
that we should not use more degrees of freedom than are 
necessary to model a phenomenon.

Moreover, classical point experiments are complete consistent 
experiments, and Copenhagen experiment are incomplete consistent 
experiments. Indeed, whenever a Copenhagen experiment assigns a 
numerical value to a quantity, the consistent experiment defined 
by the corresponding pure ensemble assigns the same value to it. 
Thus both classical mechanics and the orthodox interpretation 
of quantum mechanics are naturally embedded in the consistent 
experiment interpretation.

The logical riddles of quantum mechanics (see, e.g., 
{\sc Svozil} \cite{Svo}) find their explanation in 
the fact that most events are unsharp in a given consistent 
experiment, so that their objective reference values are no longer 
dichotomic but may take arbitrary values in $[0,1]$, by (SM).

The arithmetical riddles of quantum mechanics (see, e.g., 
{\sc Schr\"odinger} \cite{Sch}) find their explanation in the fact 
that most Hermitian quantities are unsharp in a given consistent 
experiment, so that their objective reference values are no longer 
eigenvalues but may take arbitrary values in the convex hull of the 
spectrum. 

Why then do we 'observe' only discrete values when 'measuring' 
quantities with discrete spectra? In fact one only measures related
macroscopic quantities obtained by a thermodynamic magnification 
process that forces the measurement apparatus (which interacts with
the observed system) into an equilibrium state: the dissipative 
environment selects the preferred basis in which the 'collapse of the 
wave function' happens; see the references at the end of this section.
 
Then it is claimed (on the basis of knowing
the form of the interaction between observing and observed system)
that the accurate values of the macroscopic observables obtained are in 
fact an accurate measurement of corresponding quantum 'observables'
of the observed system. But the relation is indirect, and --
as repetition shows -- such a 'measurement' is unreliable. 
One gets a reproducible result -- i.e.,  
a {\em reliable measurement of the quantum system} -- 
only by averaging a large number of events. 

Thus what is really 
(= reproducibly) observed about the quantum system is its density,
or rather the joint probability distribution of some of its quantities. 
To deduce from this density a reference value for a quantity requires 
taking an average, which is usually not in the spectrum. 
The same process also explains how joint measurements of 
complementary variables such as position and momentum are possible; 
this again yields a joint distribution whose statistical properties are 
constrained by the uncertainty relation.

The geometric riddles of quantum mechanics -- e.g., in the double 
slit experiment ({\sc Bohr} \cite{Bohr}, {\sc Wootters \& Zurek} 
\cite{WooZ}) and in EPR-experiments ({\sc Aspect} \cite{Asp}, 
{\sc Clauser \& Shimony} \cite{ClaS}) -- do not disappear. 
But they remain within the magnitudes predicted by reference radii
and uncertainties, hence require no special interpretation in the 
microscopic case. They simply demonstrate that {\em particles are 
intrinsically extended and cannot be regarded as pointlike}. 

(The extendedness of quantum particles has been mentioned in a
number of places, e.g., by {\sc Braginsky \& Khalili} \cite{BraK} 
on nonrelativistic quantum measurement theory (cf. the above quote), 
by {\sc Marolf \& Rovelli} on relativistic quantum position measurement.
For relativistic particles, extendedness is unavoidable also for 
different reasons, since it is impossible to define spacetime 
localization in a covariant manner. A lucid argument for this 
due to {\sc Haag} (unpublished) is presented in 
{\sc Keister \& Polyzou} \cite[Section 4.4]{KeiP}; cf. also
{\sc Foldy \& Wouthuisen} \cite{FolW} and 
{\sc Newton \& Wigner} \cite{NewW}.
Extendedness also shows in field theory, where particles are 
excitations of fields which are necessarily extended, except at 
special moments in time.)

When considering quantum mechanical phenomena that violate our 
geometric intuition, one should bear in mind similar violations 
of a naive geometric picture for the classical center of mass, 
Einstein's prototype example for a definite and objective property of 
macroscopic systems:
First, though it is objective, the center of mass is nevertheless 
a fictitious point, not visibly distinguished in reality; for a
nonconvex classical object it may even lie outside its boundary! 
Second, the center of mass follows a well-defined, objective path, 
though this path need not conform to the visual path of the object; 
this can be seen by pushing a drop of dark oil through a narrow, 
strongly bent glass tube. 

Compare this with an extended quantum particle squeezing itself 
through a double-slit, while its reference path goes through the 
barrier between the slits. 
(What happens during the passage? I have never seen this discussed. 
But in a detailed quantum description, the particle gets entangled 
with the double slit and loses its individuality, emerging again -- 
as an asymptotic scattering state -- unscathed only after the 
interaction has become negligible.)

Similarly, the fact that a particle hitting a screen of detectors 
excites only one of the detectors does not enforce the notion of a 
pointlike behavior; an extended flood also breaks a dam often
only at one place, that of least resistance. As the latter point
may be unpredictable but is determined by the details of the dam,
so the detector responding to the particle may simply be the one
that is slightly easier to excite. The latter is determined by the 
microstate of the detector but unpredictable, since the irreversible
magnification needed to make the measurement permanent enough to be
reliable the observed presupposes a sufficiently chaotic microstate 
(namely, according to statistical mechanics, one in local equilibrium),
whose uncertain preparation is the source of the observed randomness.

\bigskip
All the mathematical considerations above (though not all the 
illustrating comments) are independent of the measurement problem.
To investigate how measurements of classical macroscopic quantities
(i.e., expectations of quantities with small uncertainty related to a
measuring device) correlate with reference values of a microscopic
system interacting with the device requires a precise definition of a 
measuring device and of the behavior of the combined system under the
interaction (cf. the treatments in {\sc Braginsky \& Khalili} 
\cite{BraK}, {\sc Busch} et al. \cite{BusGL,BusLM}, 
{\sc Giulini} et al. \cite{GiuJK}, {\sc Mittelstaedt} \cite{Mit} 
and {\sc Peres} \cite{Per3}), and should not be considered part of 
an axiomatic foundation of physics.

\section{Dynamics} \label{dynamics}

\hfill\parbox[t]{6.5cm}{\footnotesize

{\em The lot is cast into the lap; 
but its every decision is from the {\sc LORD}.}

King Solomon, ca. 1000 B.C. \cite{Sol} 

\bigskip
{\em God does not play dice with the universe.}

Albert Einstein, 1927 A.D. \cite{Ein}
}\nopagebreak

\bigskip
In this section we discuss the most elementary aspects of the dynamics 
of (closed and isolated) physical systems. The goal is to show that 
there is no difference in the causality of (nonrelativistic) 
classical mechanics and that of quantum mechanics.

The observations about a physical system change with time. The dynamics 
of a closed and isolated physical system is conservative, and may be
described by a fixed (but system-dependent) 
one-parameter family $S_t$ ($t\in\Rz$) of {\bf automorphisms} of the 
*-algebra $\Ez$, i.e., mappings $S_t:\Ez\to\Ez$ satisfying 
(for $f,g \in \Ez$, $\alpha\in\Cz$, $s,t\in\Rz$)

(A1) 
~{$S_t(\alpha)=\alpha, ~~~ S_t(f^*)=S_t(f)^*$,}

(A2)
~{$S_t(f+g)=S_t(f)+S_t(g), ~~~ S_t(fg)=S_t(f)S_t(g)$,}

(A3)
~{$S_0(f)=f, ~~~ S_{s+t}(f)=S_s(S_t(f))$.}

In the {\bf Heisenberg picture} of the dynamics, where ensembles are
fixed and quantities change with time, $f(t):=S_t(f)$ denotes the 
time-dependent {\bf Hei\-senberg quantity} associated with 
$f$ at time $t$. Note that $f(t)$ is uniquely determined by $f(0)=f$. 
Thus the dynamics is deterministic, {\em independent of whether we 
are in a classical or in a quantum setting}.

(In contrast, nonisolated closed systems are dissipative and 
intrinsically sto\-chastic; see, e.g., 
{\sc Giulini} et al. \cite{GiuJK}.)

\begin{expls} \label{ex.classquant.dyn}
In nonrelativistic mechanics, conservative systems are described by a 
Hermitian quantity $H$, called the {\bf Hamiltonian}.

(i) In {\bf classical mechanics} 
-- cf. Example \ref{ex.classquant}(i) --,
a Poisson bracket $\{\cdot,\cdot\}$ together with $H$ defines the 
Liouville superoperator $Lf=\{f,H\}$, and the dynamics is given by 
the one-parameter group defined by
\[
S_t(f)=e^{tL}(f),
\]
corresponding to the differential equation
\lbeq{e.heisc}
\frac{df(t)}{dt}=\{f(t),H\}.
\eeq

(ii) In {\bf nonrelativistic quantum mechanics} 
-- cf. Example \ref{ex.classquant}(ii) --, 
the dynamics is given by the one-parameter group defined by
\[
S_t(f)=e^{-tH/i\hbar}fe^{tH/i\hbar},
\] 
corresponding to the {\bf Heisenberg equation}
\lbeq{e.heisq}
i\hbar\frac{df(t)}{dt}=e^{-tH/i\hbar}[f,H]e^{tH/i\hbar}=[f(t),H].
\eeq
If we write
\[
f \lp g:=\left\{\bary{rl}
-\{f,g\}             &\mbox{for a classical system,}\\
\frac{i}{\hbar}[f,g] &\mbox{for a quantum system,}
\eary\right.
\]
we find the common description
\lbeq{eq35a}
\dot f = H \lp f.
\eeq
Indeed, it is well-known that the operation $\lp$ satisfies analogous 
axioms for a commutative (classical) respective noncommutative 
(quantum) Poisson algebra; cf. \cite{Neu.qftev,Vais}.

Thus the realization of the axioms is different in the classical and 
in the quantum case, but the interpretation is identical. 

(iii) {\bf Relativistic quantum mechanics} is currently 
(for interacting systems)
developed only for scattering events in which the dynamics is 
restricted to transforming quantities of a system at $t=-\infty$ to 
those at $t=+\infty$ by means of a single automorphism $S$ given by
\[
S(f)=sfs^*,
\]
where $s$ is a unitary quantity (i.e., $ss^*=s^*s=1$), the so-called 
{\bf scattering matrix}, for which an asymptotic series in powers of
$\hbar$ is computable from quantum field theory.

\end{expls}

Of course, reference values of quantities at different times 
will generally be different. To see what happens, suppose that,
in a consistent experiment $v$ given by an ensemble, a quantity $f$ 
has reference value $v(f)$ at time $t=0$. At time $t$, the quantity 
$f$ developed into $f(t)$, with reference value 
\lbeq{e.SH}
v(f(t))=v(S_t(f))=v_t(f),
\eeq
where the time-dependent {\bf Schr\"odinger ensemble}
\lbeq{e.schro}
v_t=v\circ S_t
\eeq
is the composition of the two mappings $v$ and $S_t$.
It is easy to see that $v_t$ is again an ensemble, hence a consistent 
experiment.

Thus we may recast the dynamics in the {\bf Schr\"odinger picture}, 
where quantities are fixed and ensembles change with time. The dynamics
of the time-dependent ensembles $v_t$ is then given by \gzit{e.schro}. 
Of course, in this picture, the dynamics is deterministic, too.

\begin{expls}~\nopagebreak

(i) In {\bf classical mechanics}, \gzit{e.heisc} implies
for an consistent experiment of the form 
\[
v_t(f)=\int_{\Omega_{cl}} \rho(\omega,t)f(\omega)d\omega
\]
the {\bf Liouville equation}
\[
i\hbar\frac{d\rho(t)}{dt}=\{H,\rho(t)\}.
\]

(ii) In {\bf nonrelativistic quantum mechanics}, \gzit{e.heisq} implies
for an consistent experiment of the form 
\[
v_t(f)=\tr \rho(t)f 
\]
the {\bf von Neumann equation}
\[
i\hbar\frac{d\rho(t)}{dt}=[H,\rho(t)].
\]
\end{expls}

The common form and deterministic nature of the dynamics, independent 
of any assumption of whether the system is classical or quantum, 
implies that there is no difference in the causality of classical 
mechanics and that of quantum mechanics. Therefore, 
{\em the differences between classical mechanics and quantum mechanics 
cannot lie in an assumed intrinsic indeterminacy of quantum mechanics 
contrasted to deterministic classical mechanics}. 
The only difference between classical mechanics and quantum mechanics 
lies in the latter's lack of commutativity.

\section{The nature of physical reality} \label{reality}

\hfill\parbox[t]{10.8cm}{\footnotesize

{\em If, without in any way disturbing a system, we can predict with 
certainty (i.e., with probability equal to unity) the value of a 
physical quantity, then there exists an element of physical reality 
corresponding to this physical quantity.}

Albert Einstein, 1935 \cite{EinPR}

\bigskip
{\em Only love transcends our limitations. In contrast, our predictions
can fail, our communication can fail, and our know\-ledge can fail.
For our knowledge is patchwork, and our predictive power is 
limited. But when perfection comes, all patchwork will disappear.}

St. Paul, ca. 57 A.D. \cite{Pau}
}\nopagebreak

\bigskip
In a famous paper, {\sc Einstein, Podolsky \& Rosen} \cite{EinPR} 
introduced the criterion for elements of physical reality just cited,
and postulated that

{\em 
the following requirement for a complete theory seems to be a 
necessary one: every element of the physical reality must have a 
counterpart in the physical theory.
}

Traditionally, elements of physical reality were thought to have to
emerge in a classical framework with hidden variables.
However, to embed quantum mechanics in such a framework is impossible 
under natural hypotheses ({\sc Kochen \& Specker} \cite{KocS});
indeed, it amounts to having ensembles in which all Hermitian quantities
are sharp, and we have seen that this is impossible for quantum
systems involving a Hilbert space of dimension $4$ or more.
  
However, expectations -- the reference values of consistent 
experiments -- are such elements of physical reality: 
If one knows in an experiment $v=v_0$ all reference values with 
certainty at time $t=0$ then, since the dynamics is deterministic, one 
knows with certainty the reference values \gzit{e.SH} at any time.
In this sense, consistent experiments provide a realistic 
interpretation of quantum mechanics, consistent also with Einstein's 
intuition about the nature of reference states.

This is emphasized by the fact that, as shown, e.g., in
{\sc Marsden \& Ratiu} \cite[Example 3.2.2]{MarR},
it is possible to view the {\bf Ehrenfest theorem}
\[
\frac{d}{dt}\<f\>=\Big\<\frac{i}{\hbar}[f,g]\Big\>,
\]
which follows from \gzit{eq35a}, as a {\em classical} symplectic 
dynamics for these elements of physical reality -- namely
on the algebra whose elements (`measurables') are all sufficiently 
nice functions of expectations of arbitrary quantities.

The deterministic dynamics of the reference values appears to be
in conflict with the non-deterministic nature of {\em observed} 
reality. This conflict can be resolved by noting that
a small subsystem $M$ (a measuring apparatus) of a large quantum system
(the universe) can 'observe' of another (small or large) subsystem $S$ 
of interest only the effect of the interaction of $S$ on the state of
$M$. This limits the quality of the observations made. In particular,
we cannot not observe the objective reference values but only 
approximations, with uncertainties depending on the size of the 
observed (and the observing) system.

\bigskip
Thus we conclude that the true 'observables' of a physical system are
not the (Hermitian) quantities themselves, as in the traditional 
interpretation, but {\em expectations} of quantities. This is consistent
with the fact that probabilities (i.e., expectations of events) and 
other statistical quantities are measured (in the wide sense, 
including calculations) routinely, and as accurately 
as the law of large numbers allows. 

In particular, the long-standing question 'what is probability on the 
level of physical reality' gets the answer, 
'the result of measuring an event'. This answer
(given in a less formal context already by {\sc Margenau}
\cite[Section 13.2]{Mar}) is as accurate as the answer to any question 
relating theoretical concepts and physical reality can be. 
It gives probability an objective
interpretation precisely to the extend that objective protocols for 
measuring it are agreed upon. Here, objectivity is seen as a 
property of cultural agreement on common protocols, and not as 
something inherent in a concept. 

The subjectivity remaining lies in the question of deciding which
protocol should be used for accepting a measurement as 'correct'.
Different protocols may give different results.
Both classically and quantum mechanically, the experimental context
needed to define the protocol influences the outcome.

In particular, there is a big difference between measuring (in the 
wide sense, including calculations) an event before ({\em predicting}) 
or after ({\em analyzing}) it occurs. This is captured rigorously by 
conditional probabilities in classical probability theory, and 
nonrigorously by the 'collapse of the wave function' in quantum 
physics. Recognizing the 'collapse' as the quantum analogue of 
the change of conditional probability when new information arrives 
removes another piece of strangeness from quantum physics.

If a state is completely known then {\em everything} (all elements of
physical reality, i.e., all reference values for consistent experiments)
can be predicted with certainty, both in the quantum and in the 
classical case. But, in both cases, this only holds for an isolated 
system. In practice, physical systems (especially small, observed 
systems) are never isolated, and hence interact with the environment 
in a way we can never fully control and hence know. 
This, and only this, is what introduces unpredictability and hence 
forces us to a probabilistic description. And a change in the amount 
of information available to us to limit our uncertainty is modeled 
by conditional probability or its quantum equivalent, the 'collapse'.

To deepen the understanding reached, one would have to create a theory 
of measurement that allows to evaluate the quality of measurement 
protocols based on modeling both the observer and the observed within
a theoretical model of physical reality, and comparing the results of a
protocol executed in this model with the values it is claimed to 
measure. While this seems possible in principle, it is a much more
complex undertaking that lies outside the scope of an axiomatic
foundation of physics, though it would shed much light on the 
foundations of measurement.

\bigskip
Taking another look at the form of the Schr\"odinger dynamics
\gzit{e.SH}, we see 
that the reference values behave 
just like the particles in an ideal fluid, propagating independently 
of each other. We may therefore say that 
the Schr\"odinger dynamics describes the {\bf flow of truth}
in an objective, deterministic manner. On the other hand,
the Schr\"odinger dynamics is completely silent 
about {\em what} is true. Thus, as in mathematics, where all truth is 
relative to the logical assumptions made (namely what is considered 
true at the beginning of an argument), in theoretical physics truth 
is relative to the initial values assumed (namely what is considered 
true at the beginning of time).

In both cases, theory is about what is consistent, and not about what 
is real or true. The formalism enables us only to deduce truth from 
other assumed truths. But what is regarded as true is outside the 
formalism, may be quite subjective (unless controlled by social 
agreement on protocols for collecting and maintaining data) and may 
even turn out to be contradictory, depending on the acquired personal 
(or collective social) habits of self-critical judgment. The amount
of objectivity and reliblilty achievable depends very much on 
maintaining high and mature standards in conducting science.

What we can possibly know as true are the {\em laws} of physics, 
general relationships that appear often enough to see the underlying 
principle. But concerning the 'state of the world' (i.e., in practice, 
initial or boundary conditions) we are doomed to idealized, 
more or less inaccurate approximations of reality. 
{\sc Wigner} \cite[p.5]{Wig} expressed this by saying,
{\em the laws of nature are all conditional statements and they relate 
only to a very small part of our knowledge of the world.}

\bigskip
\section{Epilogue} 

The axiomatic foundation given here of the basic principles underlying 
theoretical physics suggests that, from a formal point of view, the 
differences between classical physics and quantum physics are only 
marginal (though in the quantum case, the lack of commutativity 
requires some care and causes deviations from classical behavior). 
In both cases, everything derives from the same assumptions simply by 
changing the realization of the axioms. 

As shown in \cite{Neu.qftev}, this view extends even to the deepest 
level of physics, making classical field theory and quantum field 
theory almost twin brothers.

\bigskip


\begin{thebibliography}{99}

\bibitem{Asp} A. Aspect,
Proposed experiment to test the nonseparability of quantum mechanics,
Phys. Rev. D 14 (1976), 1944--1951.
(Reprinted in \cite{WheZ}.)

\bibitem{Bel} J.S. Bell,
On the Einstein Podolsky Rosen paradox,
Physics 1 (1964), 195--200.
(Reprinted in \cite{WheZ}.)

\bibitem{Bel3} J.S. Bell,
On the problem of hidden variables in quantum mechanics,
Rev. Mod. Phys. 38 (1966), 447--452.
(Reprinted in \cite{WheZ}.)

\bibitem{Bel2} J.S. Bell,
Speakable and unspeakable in quantum mechanics,
Cambridge Univ. Press, Cambridge 1987.

\bibitem{Ber} H.J. Bernstein,
Simple version of the Greenberger-Horne-Zeilinger (GHZ) argument 
against local realism,
Found. Phys. 29 (1999), 521--525.

\bibitem{BirN} G. Birkhoff and J. von Neumann,
The logics of quantum mechanics,
Ann. Math. 37 (1936), 823--843.

\bibitem{Boh} D. Bohm,
A suggested interpretation of the quantum theory in terms of `hidden'
variables, I and II,
Phys. Rev. 85 (1952), 166--179.
(Reprinted in \cite{WheZ}.)

\bibitem{Bohr} N. Bohr,
Discussion with Einstein on epistemological problems in atomic physics,
pp. 200-241 in: 
P.A. Schilpp (ed.), 
Albert Einstein: Philosopher-Scientist,
The Library of Living Philosophers, Evanston 1949.
(Reprinted in \cite{WheZ}.)

\bibitem{BraK} V.B. Braginsky and F. Ya. Khalili 
Quantum measurement, 
Cambridge Univ. Press, Cambridge 1992.

\bibitem{BusGL}
P. Busch, M. Grabowski and P.J. Lahti,
Operational quantum physics,
Springer, Berlin 1995.

\bibitem{BusLM}
P. Busch, P.J. Lahti and P. Mittelstaedt,
The quantum theory of measurement, 2nd. ed.,
Springer, Berlin 1996.

\bibitem{Cir} B.S. Cirel'son, 
Quantum generalizations of Bell's inequality, 
Lett. Math. Phys. 4 (1980), 93--100.

\bibitem{ClaHS} J. F. Clauser, M.A. Horne, A. Shimony and R.A. Holt,
Proposed experiment to test local hidden-variable theories,
Phys. Rev. Lett. 23 (1969), 880--884.
(Reprinted in \cite{WheZ}.)

\bibitem{ClaS} J. F. Clauser and A. Shimony,
Bell's theorem: experimental tests and implications,
Rep. Prog. Phys. 41 (1978), 1881--1926.

\bibitem{CliK} R. Clifton and A. Kent,
Simulating quantum mechanics by non-contextual hidden variables,
Manuscript (1999). quant-ph/9908031.

\bibitem{Dav} E.B. Davies,
Quantum theory of open systems,
Academic Press, London 1976.

\bibitem{Dir} P.A.M. Dirac, 
Lectures on quantum field theory. 
Belfer Grad. School of Sci., New York 1966.

\bibitem{Dri} M. Drieschner, 
Voraussage -- Wahrscheinlichkeit -- Objekt. 
\"Uber die begrifflichen Grundlagen der Quantenmechanik.
Lecture Notes in Physics, Springer, Berlin, 1979.

\bibitem{Ebe} P.H. Eberhard,
Bell's theorem without hidden variables,
Il Nuovo Cimento 38 B (1977), 75--80.

\bibitem{Ein} A. Einstein, 
Conversation with Bohr and Ehrenfest at the Fifth Solvay conference in 
October, 1927; cf. \newl
{\small\tt http://solon.cma.univie.ac.at/\wave neum/contrib/dice.txt}
\newl
The formulation used is from a letter of September 7, 1944, 
reprinted pp. 275-276 in: 
A.P. French (ed.), 
Einstein, a centenary volume,
Harvard Univ. Press, Cambridge, Mass. 1979.

\bibitem{Ein2} A. Einstein, 
Einleitende Bemerkungen \"uber Grundbegriffe, in:
Louis de Broglie, physicien et penseur (A. George, ed.),
Albin Michel, Paris 1953.
quoted after \cite{dEs}, p. 407.

\bibitem{EinPR} A. Einstein, B. Podolsky and N. Rosen,
Can the quantum-mechanical description of physical reality be
considered complete?
Phys. Rev. 47 (1935), 777--780.
(Reprinted in \cite{WheZ}.)

\bibitem{dEs} B. d'Espagnat,
Veiled reality. An analysis of present-day quantum mechanical concepts,
Addison-Wesley, Reading, Mass., 1995.


\bibitem{Fin} T.L. Fine, 
Theory of probability; an examination of foundations. 
Acad. Press, New York 1973.

\bibitem{FolW} L.L. Foldy and S.A. Wouthuysen,
On the Dirac theory of spin 1/2 particles and its non-relativistic 
limit, 
Phys. Rev. 78 (1950), 29--36.

\bibitem{GiuJK} D. Giulini, E. Joos, C. Kiefer, J. Kupsch, 
I.-O. Stamatescu and H.D. Zeh,
Decoherence and the appearance of a classical world in quantum theory, 
Springer, Berlin 1996.

\bibitem{GreHS} D.M. Greenberger, M.A. Horne, A. Shimony and 
A. Zeilinger,
Bell's theorem without inequalities,
Amer. J. Phys. 58 (1990), 1131--1143.

\bibitem{GreHZ} D.M. Greenberger, M.A. Horne and A. Zeilinger,
Going beyond Bell's theorem,
pp. 73-76 in:
M. Kafatos (ed.), 
Bell's theorem, quantum theory, and conceptions of the universe,
Kluwer, Dordrecht 1989.

\bibitem{Hac} I. Hacking,
The emergence of probability,
Cambridge Univ. Press, Cambridge 1975.

\bibitem{Har} L. Hardy,
Quantum mechanics, local realistic theories, and Lorentz-invariant 
realistic theories,
Phys. Rev. Lett. 68 (1992), 2981--2984.

\bibitem{Har2} L. Hardy,
Nonlocality for two particles without inequalities for almost all 
entangled states,
Phys. Rev. Lett. 71 (1993), 1665--1668.


\bibitem{Hei} W. Heisenberg,
{\"U}ber den anschaulichen Inhalt der quantentheoretischen Kinematik 
und Mechanik,
Zeitschrift f. Physik 43 (1927), 172--198.
(Engl. translation: Section I.3 in \cite{WheZ}.)

\bibitem{Hig} N.J. Higham,
Accuracy and stability of numerical algorithms,
SIAM, Philadelphia 1996.


\bibitem{HofMC} R. Hoffmann, V.I. Minkin and B.K. Carpenter,
Ockham's Razor and Chemistry,
HYLE Int. J. Phil. Chem 3 (1997), 3-28.\newl
{\small \tt http://rz70.rz.uni-karlsruhe.de/~ed01/Hyle/Hyle3/hoffman.htm}

\bibitem{Hol} P.R. Holland, 
The quantum theory of motion,
Cambridge Univ. Press, Cambridge 1993.

\bibitem{HomW} D. Home and M.A.B. Whitaker,
Ensemble interpretations of quantum mechanics. A modern perspective,
Phys. Rep. 210 (1992), 223--317.

\bibitem{Isa} Isaiah 55:9,
Holy Bible, New International Version, 1984.

\bibitem{Isa2} Isaiah 65:24,
Holy Bible, New International Version, 1984.

\bibitem{Jam1} M. Jammer,
The conceptual development of quantum mechanics,
McGraw-Hill, New York 1966.

\bibitem{Jam2} M. Jammer,
The philosophy of quantum mechanics: 
the interpretations of quantum mechanics in historical perspective,
Wiley, New York 1974.

\bibitem{Jau} J.M. Jauch, 
Foundations of quantum mechanics,
Addison-Wesley, Reading, MA 1968.

\bibitem{KeiP} B.D. Keister and W.N. Polyzou,
Relativistic Hamiltonian dynamics in nuclear and particle physics,
Advances in Nuclear Physics, Vol. 20 (1991), 226--479.

\bibitem{Kla} J.R. Klauder,
Coherent states in action,
Manuscript (1997).
quant-ph/9710029

\bibitem{KocS} S. Kochen and E.P. Specker,
The problem of hidden variables in quantum mechanics,
J. Math. Mech. 17 (1967), 59--67.
(Reprinted in C.A. Hooker, ed.,
The logico-algebraic approach to quantum mechanics,
Vol. I: Historical evolution,
Reidel, Dordrecht 1975.)

\bibitem{Koh} Kohelet, Ecclesiastes 11:6,
in: Holy Bible, New International Version, 1984.

\bibitem{Kol} A.N. Kolmogorov,
Foundations of the theory of probability,
Chelsea, New York 1950.
(German original: 
Grundbegriffe der Wahr\-schein\-lich\-keits\-rech\-nung,
Sprin\-ger, Berlin 1933.)


\bibitem{Mar} H. Margenau,
The nature of physical reality,
Mc Graw-Hill, New York 1950.

\bibitem{MarRo} D. Marolf and C. Rovelli,
Relativistic quantum measurement,
Manuscript, 2002.
gr-qc/0203056

\bibitem{MarR} J.E. Marsden and T.S. Ratiu,
Introduction to Mechanics and Symmetry,
Springer, New York 1994.

\bibitem{Mer} N.D. Mermin,
Simple unified form for the major no-hidden-variables theorems,
Phys. Rev. Lett. 65 (1990), 3373--3376.

\bibitem{Mer2} N.D. Mermin,
What's wrong with these elements of reality?
Physics Today (June 1990), 9--10.

\bibitem{Mes} A. Messiah,
Quantum mechanics, Vol. 1,
North-Holland, Amsterdam 1991;
Vol. 2,
North-Holland, Amsterdam 1976.

\bibitem{Mit} P. Mittelstaedt,
The interpretation of quantum mechanics and the measurement process,
Cambridge Univ. Press, Cambridge 1998.

\bibitem{Mey} P.-A. Meyer,
Quantum probability for probabilists, 2nd. ed., 
Springer, Berlin 1995.

\bibitem{Neu.prot} A. Neumaier,
Molecular modeling of proteins and mathematical prediction of 
protein structure,
SIAM Rev. 39 (1997), 407--460.

\bibitem{Neu.surprise} A. Neumaier,
Fuzzy modeling in terms of surprise, 
Fuzzy Sets and Systems, to appear.\newl
{\tt http://www.mat.univie.ac.at/\wave neum/papers.html\#fuzzy}

\bibitem{Neu.qftev} A. Neumaier,
Quantum field theory as eigenvalue problem,
Manuscript (2003).\newl
{\tt http://www.mat.univie.ac.at/\wave neum/papers.html\#qft}

\bibitem{vNeu} J. von Neumann,
Mathematische Grundlagen der Quantenmechanik.
Springer, Berlin 1932.

\bibitem{NewW}
T.D. Newton and E.P. Wigner,
Localized states for elementary systems,
Rev. Mod. Phys 21 (1949), 400--406.

\bibitem{Ock} W. of Ockham,
Philosophical Writings, 
(ed. by P. Boehner)
Nelson, Edinburgh 1957. 

\bibitem{Par} K.R. Parthasarathy,
An introduction to quantum stochastic calculus,
Birkh\"auser, Basel 1992.

\bibitem{Pau} St. Paul, 1 Corinthian 13:8-10,
in: The New Testament. This is my paraphrase 
of a famous quote by Paul; for other renderings, see, e.g.,\newl
{\small\tt http://solon.cma.univie.ac.at/\wave neum/christ/contrib/1cor13.html}

\bibitem{Pau3} St. Paul, 1 Timothy 6:8,
in: Holy Bible, New International Version, 1984.

\bibitem{Per} A. Peres,
In compatible results of quantum measurements,
Physics Lett. A 151 (1990), 107--108.

\bibitem{Per2} A. Peres,
Two simple proofs of the Kochen-Specker theorem,
J. Phys. A: Math. Gen. 24 (1991), L175--L178.

\bibitem{Per3} A. Peres, 
Quantum theory: Concepts and methods,
Kluwer, Dordrecht 1993.

\bibitem{Pit} I. Pitowsky,
Quantum probability -- quantum logic,
Lecture Notes in Physics 321,
Springer, Berlin 1989.

\bibitem{Pla} Plato, 
Timaeus,
Hackett Publishing, Indianapolis 1999. 
The quotes (Tim. 28--29) are from the 
Project Gutenberg Etext at
\newl
{\footnotesize\tt ftp://metalab.unc.edu/pub/docs/books/gutenberg/etext98/tmeus11.txt}\newl
(the first half of the document is a commentary, then follows the 
original in English translation)


\bibitem{Ric} C.E. Rickart,
General theory of Banach algebras.
Van Nostrand, Princeton 1960.

\bibitem{Rob} H.P. Robertson,
The uncertainty principle,
Phys. Rev. 34 (1929), 163--164.
(Reprinted in \cite{WheZ}.)

\bibitem{Rou} T. Roubicek,
Relaxation in Optimization Theory and Variational Calculus, 
Walter de Gruyter, Berlin 1997.

\bibitem{Sch} E. Schr{\"o}dinger,
Die gegenw{\"a}rtige Situation in der Quantenmechanik, 
Naturwissenschaften 23 (1935), 807--812; 823--828; 844--849.
(Engl. translation: Section I.11 in \cite{WheZ}.)

\bibitem{Skl} L. Sklar, 
Physics and Chance,
Cambridge Univ. Press, Cambridge 1993.

\bibitem{Sol} King Solomon, Proverbs 16:3,
in: Holy Bible, New International Version, 1984.

\bibitem{Sto} M.H. Stone,
The theory of representations for Boolean algebras, 
Trans. Amer. Math. Soc. 40 (1936), 37--111.


\bibitem{Svo} K. Svozil,
Quantum logic. 
Springer, Berlin 1998.

\bibitem{SI} B.N. Taylor,
Guide for the Use of the International System of Units (SI),
NIST Special Publication 811, 1995 edition.\\
{\tt http://physics.nist.gov/cuu/Units/introduction.html}     

\bibitem{TitBG} W. Tittel, J. Brendel, B. Gisin, T. Herzog, H. Zbinden 
and N. Gisin,
Experimental demonstration of quantum-correlations over more than 
10 kilometers,
Phys. Rev. A 57, 3229--3232 (1998). 

\bibitem{Vai} L. Vaidman,
Variations on the theme of the Greenberger-Horne-Zeilinger proof,
Found. Phys. 29 (1999), 615--630.

\bibitem{Vais} I. Vaisman,
Lectures on the Geometry of Poisson Manifolds,
Birkh\"auser, Basel 1994.

\bibitem{Whe} J.A. Wheeler, 
Delayed-choice experiments and the Bohr-Einstein dialogue,
reprinted in: Law Without Law, pp. 182--213 of \cite {WheZ}.

\bibitem{WheZ} J.A. Wheeler and W. H. Zurek,
Quantum theory and measurement.
Princeton Univ. Press, Princeton 1983.

\bibitem{Whi} P. Whittle,
Probability via expectation, 3rd ed.,
Springer, New York 1992.
(1st ed.: Probability, Harmondsworth 1970.)

\bibitem{Wig} E.P. Wigner,
The unreasonable effectiveness of mathematics in the natural sciences,
Comm. Pure Appl. Math. 13 (1960), 1--14.

\bibitem{Wis} 
Book of Wisdom 11:20,
The Holy Bible, New Revised Standard Version, 1989.

\bibitem{WooZ} W.K. Wootters and W.H. Zurek,
Complementarity in the double-slit experiment: Quantum nonseparability 
and a quantitative statement of Bohr's principle,
Phys. Rev. D 19 (1979), 473--484.
(Reprinted in \cite{WheZ}.)

\bibitem{Zim} H.-J. Zimmermann, 
Fuzzy set theory -- and its applications, 3rd ed.,
Kluwer, Dordrecht 1996. 

\end{thebibliography}
\end{document}